 \documentclass[%
 aps,
 prb,%
 amsmath,amssymb,
 reprint,%
superscriptaddress,%
twocolumn,
]{revtex4-2}

\usepackage{physics}
\usepackage{graphicx}
\usepackage{dcolumn}
\usepackage{bm}
\usepackage{dsfont}
\usepackage[dvipsnames]{xcolor}
\usepackage{hyperref}

\begin{document}
\title{Leakage at Zero Temperature from Changes in Chemical Potential in Majorana Qubits}

\author{M.\ C.\ Goffage}
\affiliation{School of Physics, The University of New South Wales, Sydney, NSW 2052, Australia}

\author{A.\ Alase}
\affiliation{School of Physics, Centre for Engineered Quantum Systems, School of Physics, University of Sydney, New South Wales, 2006, Australia}

\author{M.\ C.\ Cassidy}
\affiliation{School of Physics, The University of New South Wales, Sydney, NSW 2052, Australia}

\author{S.\ N.\ Coppersmith}
\affiliation{School of Physics, The University of New South Wales, Sydney, NSW 2052, Australia}

\date{\today}

\begin{abstract}
Building a fault-tolerant quantum computer requires physical qubits with exceptionally low error rates. Majorana-based tetron qubits are predicted to exhibit error rates that decrease exponentially with inverse temperature and length of each topological superconducting wire in the tetron. 
In contrast to this prediction, we show that errors arising from small variations in the chemical potential grow linearly with tetron length at zero temperature. These errors stem from leakage into excited quasiparticle states, which ultimately poison Majorana modes at opposite ends of the tetron, causing errors. We further demonstrate that the dynamics of this leakage is captured by the half Landau-Zener effect, which dictates its dependence on key system parameters such as the superconducting gap, chemical potential variations, and dynamic changes in the spatial profile of Majorana modes. These results motivate further investigations into the impact of leakage on qubit performance and potential mitigation strategies.
\end{abstract}

\maketitle

The past two decades have seen significant efforts towards building a fault-tolerant quantum computer based on various platforms including superconducting circuits \cite{bravyi_2022_JAppPhys_132}, spins in semiconductors \cite{Burkard_2023_RevModPhys_95}, photonics \cite{Wang_2020_NaturePhotonics_14_84}, trapped ions \cite{bruzewicz_2019_APR_6_2}, and neutral atoms \cite{Wintersperger_2023_epjQTech_10}. However, work is still needed to close the gap between quantum error correction thresholds and the error rates of current physical qubit implementations \cite{Beverland_2022_arXiv_2211.07629}. 
Topological systems provide a promising avenue in this pursuit~\cite{Nayak_2008_RevModPhys_80_59}.
In particular, realization of topological qubits encoded in
Majorana Zero Modes (MZMs) hosted by semiconducting nanowires in proximity to 
a superconductor \cite{Kitaev_2001_Physics_uspekhi_44, Lutchyn_2010_PRL_105_01, Oreg_2010_PRL_17_02, Aghaee_2023_PRB_107_23, Alicea_2012_IOP_75, Sarma_2015_npj_01}
has drawn significant experimental efforts~\cite{Aghaee_2023_PRB_107_23, Mourik_2012_Science_336_07, Albrecht_2016_Nature_531_09, aghaee_2025_nature_638_55, aasen_2025_arXiv_2502.12252}. Crucially, such qubits are predicted to be highly robust against
changes in chemical potential, which are expected to occur during manipulation of the qubit
\cite{Karzig_2017_Phys_Rev_B_95, aghaee_2025_nature_638_55}
as well as from the charge noise \cite{Mishmash_2020_PRB_101}  
commonly encountered in solid-state devices. The goal of this work is
to establish how the rates of errors in Majorana qubits 
arising from changes in chemical potential scale with the system parameters.

Errors in Majorana qubits arise in two distinct ways. 
Let us consider the \textit{tetron} architecture, in which Majorana-based qubits are composed of
two parallel nanowires proximitized by a common superconductor~\cite{Plugge_2017_New_J_Phys_19, Karzig_2017_Phys_Rev_B_95}. The four ends of these nanowires each host an MZM. 
For tetrons of lengths comparable to the superconducting coherence length, 
the errors primarily originate from the combination of a
fluctuating chemical potential and the non-zero overlap of wavefunctions of MZMs 
on the opposite ends \cite{knapp_2018_PRB_97_12}. However, such errors are exponentially suppressed
in length due to underlying topological properties, resulting in 
coherence times growing exponentially with the length for short tetrons~\cite{knapp_2018_PRB_97_12}.

For sufficiently long tetrons, however, the errors originate primarily
via a process known as \textit{quasiparticle poisoning} (QPP). QPP originally attracted significant attention in the context of superconducting qubits \cite{Aumentado_2004_PRL_02, Lutchyn_2006_PRB_15, Martinis_2009_PRL_103_02} and has more recently been investigated in Majorana qubits.
Quasiparticles (QPs) are mobile fermionic excitations with energies above the superconducting gap. These quasiparticles may travel across the length of the tetron and \textit{poison} the MZMs through uncontrolled interactions~\cite{karzig_2021_phys_rev_let_126},
resulting in qubit errors~\cite{karzig_2021_phys_rev_let_126, Karzig_2017_Phys_Rev_B_95, knapp_2018_PRB_97_12, Alase_2024_Phys_Rev_Res_6, Rainis_2012_PRB_17, Goldstein_2011_PRB_84_09}. Unlike errors originating from non-zero overlap of MZMs, the errors originating from QPP are not suppressed by topology.
In Majorana qubits, QPP can be described as either being extrinsic \cite{knapp_2018_PRB_97_12, Rainis_2012_PRB_17, Budich_2012_PRB_85_05} or intrinsic \cite{knapp_2018_PRB_97_12} in origin. Extrinsic QPP occurs when a QP hops onto the device from the environment. This can be suppressed experimentally, for instance by implementing devices with high charging energy~\cite{Karzig_2017_Phys_Rev_B_95}. 
QPs may also be excited via intrinsic means, 
including changes in chemical potential due to charge noise \cite{Mishmash_2020_PRB_101} or gate operations \cite{Truong_2023_PhysRevB_107, Cheng_2011_PRB_84_29}, thermal excitations \cite{knapp_2018_PRB_97_12}, cosmic rays and stray radiation \cite{karzig_2021_phys_rev_let_126}. Thermal excitations are considered to be the dominant source of intrinsic QPs, 
however such excitations are exponentially 
suppressed in the ratio of band gap to the temperature and the resulting errors are linearly suppressed in the length of the nanowires \cite{knapp_2018_PRB_97_12}. 
Here we focus on the production of intrinsic QPs from changes in the chemical potential, a source that
exists even at zero temperature and which is not expected
to be suppressed by the nanowire length \cite{Mishmash_2020_PRB_101}.

\begin{figure*}
\centering
\includegraphics[width=18cm]{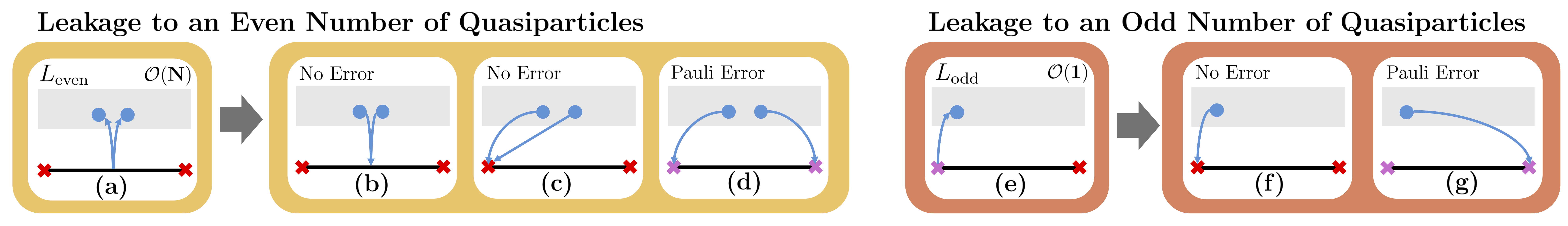}
\caption{Lowest order leakage mechanisms in a topological superconducting nanowire for an even (a) and odd (e) number of quasiparticle (QP) excitations and their possible subsequent outcomes (b-d) and (f-g), respectively. Each panel shows the energy band diagram for a single nanowire, with energy on the y-axis and position along the nanowire on the x-axis. At zero energy (actually an energy that decreases exponentially with nanowire length) the nanowire hosts Majorana Zero Modes, MZMs, (red crosses) at either end of the nanowire. Quasiparticles (blue circles) are permitted in an energy band (grey region) above the topological gap. When a QP interacts with an MZM, that MZM is poisoned (denoted by a purple cross) and its parity is flipped. Panel (a) shows two bulk quasiparticles excited which causes leakage in the even-QP-number sector ($L_{\text{even}}$). This process grows as $\mathcal{O}(N)$, where $N$ is the number of sites in the Kitaev-chain representing the nanowire. This can be either converted to no errors if the two QPs recombine (b) or if they are both absorbed by the same MZM (c). Alternatively, if the QPs are absorbed by MZMs on opposite ends of the wire, then a Pauli error occurs (d). Panel (e) shows a single QP emitted from an MZM which flips the MZM-parity and causes leakage in the odd-QP-number sector ($L_{\text{odd}}$). This process is constant in $N$ (it is of order $\mathcal{O}(1)$). The single excited QP results in no errors if it absorbed by the same MZM which it emitted it (f), and it causes a Pauli error if it absorbed by the other MZM (g).}
\label{fig:QP_process}
\end{figure*}

To estimate the QPP errors arising from chemical potential changes, we study leakage in the 
excited quasiparticle states of the tetron qubit exposed to a linear global chemical potential ramp,
while ensuring that the system stays deep within the topological phase at all times. Specifically, we consider two leakage quantities, namely leakage into excited states with even ($L_{\text{even}}$) and odd ($L_{\text{odd}}$) numbers of QPs respectively.
We find that for a tetron initialized in its even-MZM-parity ground states, $L_{\text{even}}$ grows linearly in length of the tetron, whereas $L_{\text{odd}}$ remains constant. We characterize the dependence of these leakages on the ramp rate and show that at low ramp rates the dynamics are governed by half Landau-Zener physics \cite{deGrandi_2010_Springer_75_114} between the ground subspace and the excited quasiparticle states. This is similar but distinct from previous studies which showed that full Landau-Zener physics describes Majorana-qubits subject to large local chemical potential changes that drive local topological phase transitions \cite{Bauer_2018_SciPostPhys_05_04, Truong_2023_PhysRevB_107, karzig_2015_PRB_91_04}. 

The scaling of $L_{\text{even}}$ and $L_{\text{odd}}$
we uncover can be understood from a very simple intuitive argument.
In the low-leakage regime, QP generation is dominated by two processes, namely
the generation of a pair of QPs in the bulk of the wire
(see Fig.~\ref{fig:QP_process}a), or emission of a single QP 
from an MZM (see Fig.~\ref{fig:QP_process}e). These processes contribute dominantly to $L_{\text{even}}$ and $L_{\text{odd}}$ respectively. 
For the bulk process, when a perturbation capable of exciting QPs is applied, one expects the rate of QP pair generation to grow
linearly with the length of the wire, thereby explaining the scaling of $L_{\rm even}$.
On the other hand, the rate of QP emission 
from exponentially localized MZMs is expected to remain constant in the length of the wire,
explaining why $L_{\rm odd}$ achieves a constant value for sufficiently large $N$.
Interestingly, this behavior is observed irrespective of the ramp rate
and the amplitude of change in the chemical potential. 

Note that when an MZM emits a QP ($L_{\text{odd}}$), 
the joint parity of the MZMs is flipped, which places the tetron qubit 
in a non-computational state resulting in qubit leakage~\cite{knapp_2018_PRB_97_12, Alase_2024_Phys_Rev_Res_6}. In contrast, 
generation of pairs of QPs in the bulk amounts to no qubit error instantaneously.
Nevertheless, both $L_{\text{even}}$ and $L_{\text{odd}}$ will likely have a significant impact on qubit performance. Once generated, QPs are highly mobile \cite{Alase_2024_Phys_Rev_Res_6} and may travel through the nanowire and be absorbed by the MZMs at either end \cite{karzig_2021_phys_rev_let_126}. 
It is expected that the rate of QP recombination (Fig.\ \ref{fig:QP_process}b) will be much less than QP absorption by MZMs (Fig.\ \ref{fig:QP_process}c,d) \cite{karzig_2021_phys_rev_let_126}. 
As shown in Fig.\ \ref{fig:QP_process} this can either result in a Pauli error or no errors/leakage. Assuming the QPs move diffusively and independently, we estimate that 1/3 of the QP pairs will be converted to Pauli errors.
Therefore we expect the Pauli error rate to increase with $L_{\text{even}}$ and the tetron length. These undesirable processes will likely occur after any linear chemical potential ramp, growing with system size due to the scaling of $L_{\text{even}}$. 
Our study motivates more rigorous investigations into the consequences of leakage on qubit performance, and methods for mitigating their impact. 

The rest of this paper is structured as follows. First we review 
the Kitaev tetron model of a Majorana tetron qubit, state rigorous definitions
of leakages and related quantities and briefly review the covariance matrix
method used in the numerics. We then describe the setup of the numerical
experiment and report the results including deduced scaling of the leakages. 
Finally, we discuss the implications of these results for the performance of Majorana-based qubits.

\textit{The Kitaev-Tetron Qubit.} - We model the topological superconducting wire using the discrete, tight-binding, Kitaev chain Hamiltonian \cite{Kitaev_2001_Physics_uspekhi_44}. We define fermionic annihilation $\hat{c}_j^{(\lambda)}$ and creation operators $\hat{c}_j^{(\lambda)\dagger}$ on an $N$ site 1-dimensional lattice, i.e. $j = 1, ..., N$, and where $\lambda \in \mathbb{N}^+$ indexes distinct 1-D lattices. The Kitaev chain defined on an $N$ site lattice is

\begin{eqnarray}
    \hat{H}_{\text{KC}}^{(\lambda)}(t) &&= 
        -\mu(t) \sum_{j = 1}^{N}\left(\hat{c}_j^{(\lambda) \dagger}\hat{c}_j^{(\lambda)} - \frac{1}{2}\right) + \nonumber\\
       &&\sum_{j = 1}^{N-1}
       \left(-w \hat{c}_j^{(\lambda) \dagger} \hat{c}_{j+1}^{(\lambda)}  + \Delta \hat{c}_j^{(\lambda)} \hat{c}_{j+1}^{(\lambda)} + \text{H.c.}\right),
    \label{eqn:kitaev_chain}
\end{eqnarray}

\noindent
where $\mu(t)$ is a time-dependent on-site chemical potential; $w$ is the hopping strength; $\Delta$ is the superconducting pairing strength; and H.c.\ denotes the Hermitian conjugate. Diagonalising $\hat{H}_{\text{KC}}^{(\lambda)}$ at time $t$ gives the Hamiltonian in terms of the instantaneous delocalized Bogoliubov QPs,
\begin{align}
    \hat{H}_{\text{KC}}^{(\lambda)}(t) = \sum_{k = 0}^{N-1} 
     \varepsilon_{k,t}^{(\lambda)} \hat{d}_{k,t}^{(\lambda)\dagger} \hat{d}_{k,t}^{(\lambda)} + \text{constant},
     \label{eqn:hamil_deloc_qp}
\end{align}
where $\hat{d}_{k,t}^{(\lambda)}$ and $\hat{d}_{k,t}^{(\lambda)\dagger}$ are the delocalized instantaneous QP annihilation and creation operators respectively; and $\varepsilon_k^{(\lambda)}$ are their energies. 
The Kitaev chain exhibits a topological superconducting phase when $|\mu| < 2|w|$ and $\Delta \neq 0$.
In this phase, $\varepsilon_0^{(\lambda)} \approx 0$ and the Kitaev chain has two near-degenerate ground states, and supports excited QP states ($\hat{d}_{k,t}^{(\lambda)}$ for $k \geq 1$) above the topological gap (see Appendix~\ref{sec:app_a} for further details on the spectrum of $\hat{H}_{\text{KC}}(t)$). 
Here, the Kitaev chain hosts two MZMs localized on either end of the chain at near-zero energy, given by $\hat{\gamma}_{a,t} = \hat{d}_{0,t}^{(\lambda)} + \hat{d}_{0,t}^{(\lambda) \dagger}$ and $\hat{\gamma}_{b,t} = i\left(\hat{d}_{0,t}^{(\lambda)} - \hat{d}_{0,t}^{(\lambda) \dagger}\right)$.

Due to the fermionic superselection rule, a coherent superposition cannot be formed from two near-degenerate ground states of a single topological superconducting wire, due to their opposite fermionic parity \cite{bravyi_2012_comMathPhys_316}. Therefore a single topological nanowire cannot be used as a qubit. Instead, a qubit can be formed from two topological nanowires. This is the \textit{tetron qubit} which has received significant recent attention \cite{Karzig_2017_Phys_Rev_B_95, aasen_2025_arXiv_2502.12252}. 
We model the tetron qubit as two uncoupled Kitaev chains, 
\begin{eqnarray}
    \hat{H}(t) = \hat{H}_{KC}^{(1)}(t) + \hat{H}_{KC}^{(2)}(t),
    \label{eqn:tetron_Hamilt}
\end{eqnarray}
where in our calculation the Kitaev-chain parameters $\mu(t),\, \Delta,\, t$ are identical for both Kitaev chains. The tetron qubit has four MZMs,
\begin{align}
    \begin{matrix}
        \hat{\gamma}_{1,t} = \hat{d}_{0,t}^{(1)} + \hat{d}_{0,t}^{(1) \dagger} &&
        \hat{\gamma}_{2,t} = i\left(\hat{d}_{0,t}^{(1)} - \hat{d}_{0,t}^{(1) \dagger}\right) \\
        \hat{\gamma}_{3,t} = \hat{d}_{0,t}^{(2)} + \hat{d}_{0,t}^{(2) \dagger}  &&
        \hat{\gamma}_{4,t} = i\left(\hat{d}_{0,t}^{(2)} - \hat{d}_{0,t}^{(2) \dagger} \right),         
    \end{matrix}
\end{align} 
which gives four near-degenerate ground states. We define the computational basis states as those with even MZM parity. Let $\ket{\Omega_t}$ be the ground state that is annihilated by all quasiparticles, i.e.\ $\hat{d}_{k,t}^{(\lambda)}\ket{\Omega_t} = 0,~ \forall k,~ \lambda \in \{1,2\}$. Then the ground states with even parity are $\ket{0_t} = \ket{\Omega_t}$ and $\ket{1_t} \propto \hat{d}_{k,t}^{(1) \dagger} \hat{d}_{k,t}^{(2) \dagger} \ket{\Omega_t}$, which respectively have MZM-parity on each chain of $\ket{0_t} = \ket{\text{even}, \text{even}}$ and $\ket{1_t} = \ket{\text{odd}, \text{odd}}$. Both states have total even-MZM-parity, i.e\ they satisfy $\bra{0} \hat{P}_t \ket{0} = \bra{1} \hat{P}_t \ket{1} = +1$, where $\hat{P}_t$ is the instantaneous MZM-parity operator for both Kitaev chains, 
\begin{eqnarray}
    \hat{P}_t = (i \hat{\gamma}_{1,t} \hat{\gamma}_{2,t})(i \hat{\gamma}_{3,t} \hat{\gamma}_{4,t})
            = -\hat{\gamma}_{1,t} \hat{\gamma}_{2,t} \hat{\gamma}_{3,t} \hat{\gamma}_{4,t}.
    \label{eqn:parity_operator}
\end{eqnarray}

\textit{Defining Leakage Quantities for the Tetron Qubit.} Here we define the two leakage quantities $L_{\text{even}}$ and $L_{\text{odd}}$ which are respectively the leakages into states with even and odd numbers of QPs. We assume that the tetron is initialized in a superposition of the computational states with zero excited quasiparticles, $\ket{0_t}$ and $\ket{1_t}$. Then $L_{\text{odd}}(t)$ is simply,
\begin{eqnarray}
    L_{\text{odd}}(t) = \frac{1}{2}\left(1 - \bra{\Psi(t)} \hat{P}_t \ket{\Psi_t}\right),
    \label{eqn:Lp_definition}
\end{eqnarray}
where $\ket{\Psi(t)}$ is the Kitaev-tetron state at time $t$. To write down the expression for $L_{\text{even}}$, we first define the \textit{ground subspace leakage} $L_g$, which is the total leakage out of the ground states with even MZM parity, 
\begin{eqnarray}
    L_g(t) &&= L_{\text{even}}(t) + L_{\text{odd}}(t) \\
    &&= 1 - \left|\bra{0_t}\ket{\Psi(t)} \right|^2 - \left|\bra{1_t}\ket{\Psi(t)} \right|^2.
\end{eqnarray}
Then the leakage into the even-QP-number states is
\begin{eqnarray}
    L_{\text{even}}(t) = L_g(t) - L_{\text{odd}}(t).
\end{eqnarray}
Unlike dephasing errors due to finite overlap of MZMs, these leakage quantities and the errors arising as a direct consequence of them (see Fig.\ \ref{fig:QP_process}) are not topologically suppressed.

\textit{The Kitaev-Tetron under a Linear Chemical Potential Ramp.} We consider a simple scenario where the Kitaev-tetron is exposed to a chemical potential $\mu(t)$ which is ramped linearly from an initial value $\mu_{\rm in}$ to the final value $\mu_{\rm fin}$. Similar to Ref.~\cite{Mishmash_2020_PRB_101}, we initialize the Kitaev-tetron at $t = 0$ in the $\ket{+}$ state,  
\begin{eqnarray}
    \ket{\Psi(0)} = \ket{+} = \frac{1}{\sqrt{2}}\left(\ket{0_0} + \ket{1_0}\right).
    \label{eqn:init_state}
\end{eqnarray}
We then subject the Kitaev-tetron to a linearly ramped chemical potential, on both Kitaev chains,
\begin{eqnarray}
    \mu(t) = v t, \quad 0 \leq t < T,
    \label{eqn:chem_pot_ramp}
\end{eqnarray}
\noindent
where $v$ is the ramp rate. We choose the initial chemical potential to be $\mu(0) = 0$ and we consider final chemical potentials $\mu(T) = \mu_{\text{fin}}$ deep within the topological phase (i.e. $\mu_{\text{fin}} \leq 2 |w| /10$). This is a simple tractable model that has commonalities with chemical potential changes arising from gate operations as well as the high-frequency component nature of 1/f charge noise. 
We numerically compute the leakage quantities at the end of the chemical potential ramp, $L_\text{even}(T)$ and $L_{\text{odd}}(T)$, and study their dependence on the ramp rate $v$ and chain length $N$.
These numerics are made tractable by the use of the covariance matrix method, which we review next.

\textit{The Covariance Matrix Method.} The tetron qubit has $2N$ lattice sites and therefore has a Fock space with dimension $2^{2N} \times 2^{2N}$. This Fock space is prohibitively large for numerical time evolution calculations even at modest chain lengths. However, as long as the 
initial state is a fermionic Gaussian state and time evolutions are generated
by quadratic Hamiltonians, methods leveraging the algebra of quadratic operators 
can be employed to speed up the computation 
exponentially~\cite{Mascot_2023_PRL_131,Surace_2022_SciPost_Lec, Mishmash_2020_PRB_101}.
In this work, we use the covariance matrix 
method~\cite{Surace_2022_SciPost_Lec, Mishmash_2020_PRB_101} for this purpose.
This method is based on the fact that a fermionic Gaussian state is uniquely specified
by a matrix of expectation values of quadratic observables called
the \textit{covariance matrix}; the expectation values of quartic and higher-weight
observables can be deduced using Wick's theorem.
The $4N \times 4N$ covariance matrix for the initial state of the tetron is time evolved
 at discrete time-steps under $\hat{H}(t)$. 
 Overlaps of quantum states and expectation values of any product of an even number of fermionic operators can be extracted from this matrix using formulae based on Wick's theorem. 
 This allows us to numerically calculate $L_\text{even}(t)$ and $L_\text{odd}(t)$,
 as explained in Appendix~\ref{sec:app_b}. 

\begin{figure}[b]
\includegraphics[width=8.5cm]{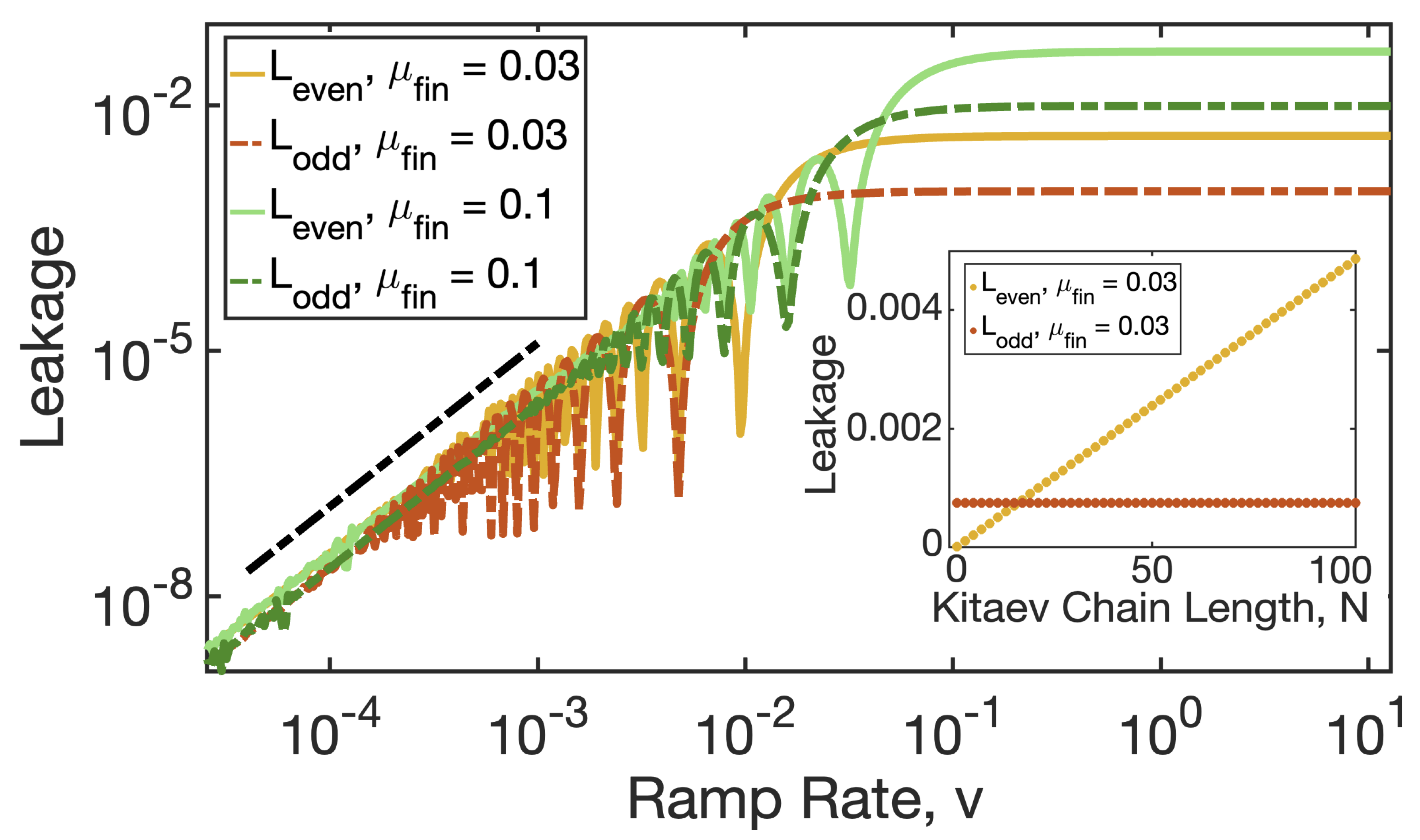}
\caption{Leakage into the sectors with an even/odd number of quasiparticles (QPs) $L_{\text{even/odd}}$ after a linear chemical potential ramp (at $t=T$) for a Kitaev-tetron qubit with initial chemical potential $\mu_\text{in} = 0$ and with hopping $w$ and pairing $\Delta$ of $w = \Delta = 0.5$. Results were calculated using the numerical covariance matrix method as detailed in the Appendix~\ref{sec:app_a}. The main figure shows $L_\text{even}$ and $L_\text{odd}$ versus the ramp rate for a Kitaev chain length of $N = 40$ and with final chemical potentials of $\mu_{\text{fin}} = 0.03$ and $\mu_{\text{fin}} = 0.10$. The black dashed reference line has gradient $2$ on the log-log scale. The inset presents $L_\text{even}$ and $L_\text{odd}$ versus the Kitaev chain length for even $N \in [2, 100]$ for a ramp rate of $v = 2\times10^{-2}$ and final chemical potential of $\mu_{\text{fin}} = 0.03$.}
\label{fig:leak_scale}
\end{figure}

{\textit{Results.} Here we detail the dependence of the leakages into the sectors with even ($L_{\text{even}}$) and odd ($L_{\text{odd}}$) numbers of quasiparticles for a Kitaev-tetron qubit following a linear chemical potential ramp, on the chemical potential ramp rate $v$ and the number of sites in a single Kitaev chain $N$.
Our numerics reveal that for sufficiently large $N$, 
$L_{\text{even}}$ scales linearly with $N$, whereas $L_{\text{odd}}$ remains 
constant in $N$ (see the inset in Fig.\ \ref{fig:leak_scale}). 
This behavior holds for all values of the ramp rate $v$ and final chemical potential $\mu_{\rm fin} \approx \mu_{\rm in}$. 
The linear scaling $L_\text{even}$ in $N$ is consistent with a 
roughly constant (independent of $N$) probability of exciting QPs along the Kitaev chains. 

The scaling of leakage with $v$ provides further insight.
Both $L_\text{odd}$ and $L_{\text{even}}$ exhibit two limiting regimes as a function of $v$,
namely near-adiabatic and sudden regimes. 
At low ramp rates $v$, both leakage quantities are in the \textit{near-adiabatic} regime where they scale as $L_\text{even/odd} \propto v^2$. This dependence can be understood 
as the cumulative result of half Landau-Zener dynamics~\cite{deGrandi_2010_Springer_75_114} 
between one of the ground states and the excited quasiparticle states. 
Further, $L_\text{even}$ and $L_\text{odd}$ oscillate at a frequency 
that is a function of the dynamic-phase accumulated by the leakage energy level, which is also predicted by this theory (see Appendix~\ref{sec:app_d}).   

At high ramp rates, both leakage quantities are in the \textit{sudden} regime, 
where they approach respective constant values, which we denote by $L_{\rm odd}(\infty)$
and $L_{\rm even}(\infty)$ respectively. 
In Appendix~\ref{sec:app_c} we compute $L_{\text{even}}(\infty)$ analytically for a system with periodic boundary conditions (which is well-justified because $L_{\text{even}}(\infty)$ is proportional to the tetron length), obtaining results that agree well with numerical results shown in Fig.~\ref{fig:leak_scale}.
Interestingly, $L_{\text{odd}}(\infty)$ 
can be approximated exponentially well by 
\begin{equation}
    L_{\rm odd}(\infty) = 1-\prod_{l}\expval{\hat{\gamma}_{l,0}|\hat{\gamma}_{l,0^+}},
\end{equation}
where $\hat{\gamma}_{l,0^+}$ denote MZMs after the sudden quench and 
$\expval{\hat{\gamma}_{l,0}|\hat{\gamma}_{l,0^+}}$ is the overlap of the 
$l$'th MZM wavefunctions before and after the quench (see
Appendix~\ref{sec:app_c}). The deviation of these overlaps away from unity 
can be interpreted as the poisoning of individual MZMs~\cite{Goldstein_2011_PRB_84_09} 
as a result of the sudden quench.
$L_{\text{even}}$ can similarly be approximated in terms of the overlaps of
bulk QP wavefunctions before and after the quench, which further reveals
quadratic dependence on the change in chemical potential.
Numerics for finite $v$ in the sudden regime show that the leakages (at $t = T$, the end of the chemical potential ramp) scale as
\begin{align}
    L_{\rm odd}(v) &= L_{\rm odd}(\infty) - \frac{k_{\rm odd}}{v^2},\nonumber\\
    L_{\rm even}(v) &= L_{\rm even}(\infty) - \frac{k_{\rm even}}{v^2},
\end{align}
where $k_{\rm odd}, ~k_{\rm even} > 0$. 

\textit{Discussion.} The results presented above demonstrate that a ramp of the chemical potential induces leakage into the sector with even numbers of QPs ($L_{\text{even}}$) that grows linearly with the tetron length even at zero temperature. Here we discuss the implications of this leakage for qubit performance. While the excited quasiparticles (QPs) themselves do not change the Majorana Zero Mode (MZM) parity and so their presence is not necessarily problematic for qubit performance \cite{karzig_2021_phys_rev_let_126, knapp_2018_PRB_97_12, Rainis_2012_PRB_17, Alase_2024_Phys_Rev_Res_6}, because the QPs are expected to be mobile and interact strongly with the MZMs at the boundaries of the wires \cite{karzig_2021_phys_rev_let_126}, this leakage process could lead to qubit errors. This is particularly likely since the relaxation of QP pairs is very slow \cite{karzig_2021_phys_rev_let_126} and the typical velocity of the QPs is high, on the order of the Fermi velocity~\cite{Alase_2024_Phys_Rev_Res_6}. As such the QPs are highly likely to travel through the wire and interact with the MZMs. 

If a pair of QPs is excited, then the excitation-trapping process does not result in an error if the QPs are absorbed by the same MZM.  However, if the two QPs are trapped by MZMs on opposite sides of the wire, then the excitation-trapping process gives rise to a Pauli error (see Fig.\ \ref{fig:QP_process}). If one assumes that the QPs move diffusively \cite{karzig_2021_phys_rev_let_126} and independently, then the probability of an error resulting from the excitation of a QP pair is given by the probability that they are absorbed by MZMs on opposite sides of the wire.
We estimate this probability by assuming the QPs in a pair are generated at the same lattice site and then follow independent random walks \cite{Doyle_1984_AMS_22} until each reaches a chain end. Averaging the probability that the two QPs reach opposite ends, for QPs generated over all lattice sites and taking the long chain limit, gives a Pauli error probability of $1/3$ (see Appendix~\ref{sec:app_e}). If instead the QPs move ballistically, one would expect the Pauli error probability to be even higher due to the opposite momenta of the constituents of the QPs.

Thus, it is possible that the excitation of QPs will result in a rate of Pauli errors that increases with system size for a real tetron qubit following a chemical potential ramp.
This result highlights the need for improved understanding of QPs in Majorana wires and also for further improvements of the design of methods for protecting against QP poisoning, a topic on which some work in the literature has already been performed~\cite{bravyi_2012_comMathPhys_316, Alase_2024_Phys_Rev_Res_6, knapp_2018_Quantum_88}.

As opposed to $L_{\text{even}}$, the Pauli error rate arising from $L_{\text{odd}}$ (leakage into states with odd numbers of QPs) is expected to decrease as the nanowire length is increased.
As shown above, the rate at which QPs are generated via $L_{\text{odd}}$ does not grow with system size.
In addition, QPs arising from $L_{\text{odd}}$ are generated near an end of the wire, and if the QP is absorbed at the MZM at the nearest end, then no decoherence results.
Simple estimates using the random walk results in Ref. \cite{Doyle_1984_AMS_22} yield a qubit error rate that decreases as the inverse of the wire length.

Since our results are based on the Kitaev chain \cite{Kitaev_2001_Physics_uspekhi_44}, our results are not only applicable to MZMs arising in hybrid semiconductor-superconductor systems \cite{Mourik_2012_Science_336_07, aghaee_2025_nature_638_55, Aghaee_2023_PRB_107_23}; but are relevant to any architecture that is described by the Kitaev chain, including MZMs in quantum dot-superconductor arrays \cite{dvir_2023_nature_614_50, leijnse_2012_PRB_86_28, Tsintzis_2024_PRXQuantum_5_23} and cold atom arrays \cite{Liang_2011_APS_106_02}. 

\textit{Conclusions.} In summary, we study the leakage out of the ground-subspace of the Kitaev-tetron qubit resulting from a small variation of the global chemical potential at a constant rate. More specifically, we study the leakage into states with even ($L_{\text{even}}$) and odd ($L_{\text{odd}}$) numbers of QPs respectively. For tetrons with chain length $N$ much larger than the localization length of the MZMs, we show that $L_{\text{even}}$ grows linearly with $N$, whereas $L_{\text{odd}}$ remains constant in $N$. We prove analytically that these scaling laws hold throughout the topological phase. We argue that the scaling of $L_{\text{even}}$ is particularly concerning as it will likely translate to qubit Pauli error rates growing with $N$. As chemical potential changes are unavoidable during qubit operations, it is crucial to investigate the extent to which the mechanism of errors studied in this work alters previous predictions of error rates being exponentially suppressed in $N$.
 
We further uncover a complete physical picture of these leakage processes for adiabatic and sudden changes in the chemical potential respectively. In the adiabatic regime, we show that $L_{\text{odd}}$ ($L_{\text{even}}$) is a cumulative result of half Landau-Zener transitions between the ground and the singly (doubly) excited states. In the sudden regime, the behavior of $L_{\text{odd}}$ ($L_{\text{even}}$) is dictated by the overlap between the initial and the final MZM (bulk QP) wavefunctions. These results allow us to establish the dependence of leakage on the rate of change of chemical potential and band gap, thereby informing strategies to prevent or correct QP-poisoning errors arising from global chemical potential changes.

\begin{acknowledgements}
We acknowledge useful conversations with Eric Bach, Salini Karuvade, and Eric Mascot.
Work at UNSW was supported by the Australian Research Council, Project No.\ DP210101608 and by the Australian Research Council Centre of Excellence in Future Low-Energy Electronics Technologies (FLEET), project no.\ CE170100039, funded by the Australian government, and by Google Asia Pacific Pte. Ltd.  M.C.G acknowledges additional support from the Sydney Quantum Academy.
A.A.\ acknowledges support by the Australian Research Council Centre of Excellence for Engineered Quantum Systems (Grant No. CE170100009). M.C.C.\ acknowledges support from a UNSW Scientia Fellowship and an Australian Research Council Discovery Early Career Research Fellowship (Grant No.\ DE240100590).
\end{acknowledgements}

\appendix
\newpage
\onecolumngrid

\section{Leakage in the Majorana Tetron Qubit} 
\label{sec:app_a}
In this section of the appendices we provide further details concerning the background of the model and numerical method detailed in the main text. We begin by reviewing the Bogoliubov-de Gennes formalism, a commonly used formalism for solving the dynamics of quadratic fermionic Hamiltonians, with a focus on the of the Kitaev-tetron Hamiltonian. We then give further remarks concerning some subtitles of encoding a qubit in the Majorana Zero Modes (MZMs) of the tetron and additionally show how to solve the spectrum of the Kitaev chain by imposing periodic boundary conditions. Following this we review Wannier quasiparticles, which allow us to formally define spatially localized quasiparticles. This section of the appendices then concludes with some remarks on the definitions of the leakage quantities $L_{\rm even}$ and $L_{\rm odd}$ employed in the main text.

\subsection{The Bogoliubov-de Gennes Formalism for the Tetron Qubit}

In the main text, we model the tetron qubit~\cite{Karzig_2017_Phys_Rev_B_95} using two uncoupled Kitaev Hamiltonians~\cite{Kitaev_2001_Physics_uspekhi_44}. The Kitaev chain $\hat{H}_{\text{KC}}^{(\lambda)}(t)$, with a time-dependent chemical potential, defined on an one-dimensional (1D) lattice, with $N$ sites is
\begin{eqnarray}
    \hat{H}_{\text{KC}}^{(\lambda)}(t) &&= 
        -\mu(t) \sum_{j = 1}^{N}\left(\hat{c}_j^{(\lambda) \dagger}\hat{c}_j^{(\lambda)} - \frac{1}{2}\right) + 
       \sum_{j = 1}^{N-1}
       \left(-w \hat{c}_j^{(\lambda) \dagger} \hat{c}_{j+1}^{(\lambda)}  + \Delta \hat{c}_j^{(\lambda)} \hat{c}_{j+1}^{(\lambda)} + \text{H.c.}\right),
    \label{eqn:kitaev_chain_repro}
\end{eqnarray}
\noindent
where this equation is reproduced from Eq. (1) in the main text. We use $\lambda \in \{1,2\}$ to index the distinct 1D lattices of the two Kitaev chains; $\hat{c}_j^{(\lambda)}$ and $\hat{c}_j^{(\lambda) \dagger }$ are respectively the fermionic annihilation and creation operators on site $j$ of chain $\lambda$, which satisfy the fermionic canonical anticommutation rules (CARs) $\{\hat{c}_i^{(\lambda)}, \hat{c}_j^{(\zeta) \dagger} \} = \delta_{i,j} \delta_{\lambda,\zeta}$ and $\{\hat{c}_i^{(\lambda)}, \hat{c}_j^{(\zeta)} \} = 0$, where $\delta_{i,j}$ is the kronecker-delta and $\{A, B\} = AB + BA$. As written in the main text, the tetron Hamiltonian is then,
\begin{eqnarray}
    \hat{H}(t) = \hat{H}_{KC}^{(1)}(t) + \hat{H}_{KC}^{(2)}(t),
    \label{eqn:tetron_Hamilt_app}
\end{eqnarray}
\noindent
which is defined on two separate 1D lattices or ``chains", giving a total of $2N$ sites.
As a consequence of the fermionic CARs, the mutually commuting number operators on each lattice site $\{ \hat{c}_j^{(\lambda) \dagger} \hat{c}_j^{(\lambda)}\}$ each have eigenvalues $\{0,1\}$. This gives a Hilbert space with dimension $2^{2N}$ \cite{Nielsen_2005_UQ}, known as the Fock Space $\mathcal{F}_{2N}$. 
The Bogoliubov-de Gennes (BdG) formalism offers a useful framework for treating quadratic fermionic operators, such as $\hat{H}(t)$ in $\mathcal{F}_{2N}$ \cite{Surace_2022_SciPost_Lec, Alicea_2012_IOP_75, Alase_2024_Phys_Rev_Res_6}. In this formalism, we define the BdG Hamiltonian $H(t)$ (which is a $4N \times 4N$ matrix, as opposed to an operator represented by a $2^{2N} \times 2^{2N}$ matrix in $\mathcal{F}_{2N}$) corresponding to $\hat{H}(t)$, according to
\begin{eqnarray}
    \hat{H}(t) = \frac{1}{2} \mathbf{\hat{c}}^\dagger H(t) \mathbf{\hat{c}},
    \label{eqn:bdg_form}
\end{eqnarray}
where 
\begin{eqnarray}
    \hat{\mathbf{c}} = \left(
    \hat{c}_1^{(1)}, \hat{c}_2^{(1)}, \cdots, \hat{c}_N^{(1)}, \hat{c}_1^{(1) \dagger}, \hat{c}_2^{(1) \dagger}, \cdots, \hat{c}_N^{(1)\dagger},
    \hat{c}_1^{(2)}, \hat{c}_2^{(2)}, \cdots, \hat{c}_N^{(2)},  
    \hat{c}_1^{(2)\dagger}, \hat{c}_2^{(2)\dagger}, \cdots, \hat{c}_N^{(2)\dagger} 
    \right)^{\text{T}}.
    \label{eqn:c_vec}
\end{eqnarray}
Let us index the elements of the matrix $H(t)$ as $H_{\text{tetron}~ (jm \lambda, j'm' \lambda')}(t)$ where $j, j' \in \{1, \cdots, N\}$ indexes the sites of a single 1D lattice; 
 $m, m' \in \{1,2\}$ indexes the creation and annihilation operators; and
 $\lambda, \lambda' \in \{1,2\}$ indexes the two chains. Of course, since the tetron consists of two uncoupled Kitaev chains on separate lattices, $H_{\text{tetron}~({jm 2, j'm' 1})}(t) = H_{\text{tetron}~({jm 1, j'm' 2})}(t) = 0$. The BdG Hamiltonian $H(t)$ represents the action of the commutator between $\hat{H}(t)$ and linear fermionic operators \cite{Alase_2024_Phys_Rev_Res_6}, 
\begin{eqnarray}
    \left[\hat{H}(t), ~\hat{c}_j^{(\lambda) \dagger} \right] &&= 
    \sum_{\lambda' = 1}^2 \sum_{j' = 1}^N 
    H_{\text{tetron}~(j1 \lambda, j' 1 \lambda')}(t) \hat{c}_{j'}^{(\lambda') \dagger} + H_{\text{tetron}~(j1 \lambda, j'2 \lambda')}(t) \hat{c}_{j'}^{(\lambda')}  \nonumber \\
    \left[ \hat{H}(t), ~\hat{c}_j^{(\lambda)} \right]  &&= 
    \sum_{\lambda' = 1}^2 \sum_{j' = 1}^N 
    H_{\text{tetron}~(j 2 \lambda, j' 1 \lambda')}(t) \hat{c}_{j'}^{(\lambda') \dagger} + H_{\text{tetron}~(j 2 \lambda, j'2 \lambda')}(t) \hat{c}_{j'}^{(\lambda')}.
\end{eqnarray}
Let us define the $\mathcal{H}_{BdG} = \text{span}\{\hat{c}_j^{(\lambda)},~\hat{c}_j^{(\lambda)\dagger} \}$ as the vector space of linear operators in the Fock space $\mathcal{F}_{2N}$. In the literature there are various conventions for denoting the eigenvectors of $\mathcal{F}_{2N}$, we keep our approach closest in spirit to Ref. \cite{Alase_2024_Phys_Rev_Res_6} and denote the basis vectors of $\mathcal{H}_{BdG}$ as $\{\ket{j, 1, \lambda} = \hat{c}_{j}^{(\lambda) \dagger}, \ket{j, 2, \lambda} = \hat{c}_{j}^{(\lambda)}\}$. Throughout the appendices we use small English and Greek letters in bras and kets to denote vectors in $\mathcal{H}_{BdG}$, and we use capital English and Greek letters to denote vectors in $\mathcal{F}_{2N}$. Importantly, due to the structure of Eq.~(\ref{eqn:bdg_form}), BdG matrices exhibit particle-hole symmetry
\begin{eqnarray}
    (\tau_x \kappa)H(t)(\tau_x \kappa)^{-1} = -H(t),
    \label{eqn:p_h_sym}
\end{eqnarray}
where $\kappa$ denotes complex conjugation and $\tau_x$ denotes particle-hole exchange, i.e. $\tau_x \ket{j,1, \lambda} = \ket{j,2, \lambda}$ and $\tau_x \ket{j, 2, \lambda} = \ket{j, 1, \lambda}$. A direct consequence of particle-hole symmetry is that for any BdG eigenvector $\ket{d_{k,t}^{(\lambda)}}$ of $H(t)$ with energy $\varepsilon_{k,t}^{(\lambda)} > 0$, there also exists an eigenvector $\ket{d_{k,t}^{(\lambda) \dagger}} \equiv \tau_x \kappa \ket{d_{k,t}^{(\lambda)}}$ with energy $-\varepsilon_{k,t}^{(\lambda)}$. For a given Kitaev chain ${H}_{\text{KC}}^{(\lambda)}(t)$ there are $N$ such eigenvectors $\ket{d_{k,t}^{(\lambda)}}$ and therefore $N$ additional eigenvectors $\tau_x \kappa \ket{d_{k,t}^{(\lambda)}}$. We index these eigenvectors with $k \in \{0, 1, \cdots, N-1\}$ with the eigenvalues $\varepsilon_{k,t}^{(\lambda)}$ in non-decreasing order. If ${H}_{\text{KC}}^{(\lambda)}(t)$ has $s_{0}^{(\lambda)}$ degenerate eigenvectors at zero energy, then $s_{0}^{(\lambda)}$ is an even number and $s_{0}^{(\lambda)}/2$ of these eigenvectors are included within the set $\left\{\ket{d_{k,t}^{(\lambda)}}\right\}$. All $4N$ eigenvectors of $H(t)$ are orthogonal and can be normalised to be orthonormal. Therefore, we can form an orthonormal basis in $\mathcal{H}_{\text{BdG}}$ from the $4N$ such vectors $\left\{\ket{d_{k,t}^{(\lambda)}},~ \tau_x \kappa \ket{d_{k,t}^{(\lambda)}}\right\}$ with $\lambda \in \{1, 2\}$. We can now write the diagonalized BdG tetron Hamiltonian in terms of its orthonormal instantaneous eigenvectors as, 
\begin{eqnarray}
    H(t) = \sum_{\lambda =1 }^2 \sum_{k = 0}^{N-1} 
    \varepsilon_{k,t}^{(\lambda)} \left( \ket{d_{k,t}^{(\lambda)}} \bra{d_{k,t}^{(\lambda)}}
    -  \tau_x \kappa \ket{d_{k,t}^{(\lambda)}} \bra{d_{k,t}^{(\lambda)}} \kappa \tau_x \right).
    \label{eqn:bdg_hamil_diagonalized}
\end{eqnarray}
This expression written in terms of $\mathcal{H}_{\text{BdG}}$ is equivalent to the diagonalized form of $\hat{H}(t)$ written in terms of fermionic operators in $\mathcal{F}_{2N}$ given in Eq. (2) in the main text, for a single Kitaev chain, and reproduced here for the full tetron as,
\begin{eqnarray}
    \hat{H}^{(\lambda)}(t)~ &&=  
        \frac{1}{2} \sum_{\lambda =1 }^2 \sum_{k = 1}^{N} 
     \varepsilon_{k,t}^{(\lambda)} \left( 
     \hat{d}_{k,t}^{(\lambda)\dagger} \hat{d}_{k,t}^{(\lambda)}  
    - \hat{d}_{k,t}^{(\lambda)} \hat{d}_{k,t}^{(\lambda) \dagger} 
     \right)
     \nonumber \\
     &&= \sum_{\lambda =1 }^2 \sum_{i = 1}^{N} 
     \varepsilon_{k,t}^{(\lambda)} \hat{d}_{k,t}^{(\lambda)\dagger} 
     \hat{d}_{k,t}^{(\lambda)} + E_{0,t},
     \label{eqn:hamil_deloc_qp_rep}
\end{eqnarray}
where $E_{0,t}$ is the ground state energy. 

\subsection{Encoding a Qubit in the Four Majorana Zero Modes of the Tetron}

As discussed in the main text, the Kitaev chain exhibits a topological phase for $|\mu| < 2|w|$ and $\Delta \neq 0$, which is characterized by the support of a near-zero energy fermionic mode $\hat{d}_{0,t}^{(\lambda)}$ with energy $\varepsilon_{0,t}^{(\lambda)} \approx 0$. The wavefunction of this fermionic mode is delocalized with weight exponentially localized on either end of the chain. It is convenient to decompose this mode into two Majorana modes that are exponentially localized on each end of the chain. Since these Majorana modes are at near-zero energy, they are referred to as Majorana Zero Modes (MZMs). Therefore, for the two Kitaev chains of the tetron we have four MZMs, associated with a delocalized fermionic mode on each chain, which as stated in the main text are
\begin{align}
    \begin{matrix}
        \hat{\gamma}_{1,t} = \hat{d}_{0,t}^{(1)} + \hat{d}_{0,t}^{(1) \dagger} &&
        \hat{\gamma}_{2,t} = i\left(\hat{d}_{0,t}^{(1)} - \hat{d}_{0,t}^{(1) \dagger}\right) \\
        \hat{\gamma}_{3,t} = \hat{d}_{0,t}^{(2)} + \hat{d}_{0,t}^{(2) \dagger}  &&
        \hat{\gamma}_{4,t} = i\left(\hat{d}_{0,t}^{(2)} - \hat{d}_{0,t}^{(2) \dagger} \right),         
    \end{matrix}
\end{align} 

As stated in the main text, it is conventional to label the even-MZM-parity ground states as $\ket{0_t} = \ket{\Omega_t}$, where $\ket{\Omega_t}$ is the vacuum of all quasiparticles, i.e.\ $d_{i,t}^{(\lambda)}\ket{\Omega_t} = 0, \forall i, \lambda \in \{1,2\}$, and $\ket{1_t} \propto d_{i,t}^{(1) \dagger} d_{i,t}^{(2) \dagger} \ket{\Omega_t}$. These states have MZM-parity on each chain of $\ket{0_t} = \ket{\text{even}, \text{even}}$ and $\ket{1_t} = \ket{\text{odd}, \text{odd}}$, they have no excited fermionic modes present, and therefore have even total fermionic parity. This means that superpositions of these states are allowed by the fermionic superselection rule. While a qubit can be formed strictly out of these states, it is more appropriate to define the qubit encoded in the four MZMs according to the expectation values of the Pauli operators \cite{Alase_2024_Phys_Rev_Res_6}, which are
\begin{eqnarray}
    \hat{Z_t} = -i \hat{\gamma}_{1,t} \hat{\gamma}_{2,t}, ~~~~ 
    \hat{X_t} = -i \hat{\gamma}_{1,t} \hat{\gamma}_{3,t},~~~~
    \hat{Y_t} = -i \hat{\gamma}_{2,t} \hat{\gamma}_{3,t}.
    \label{eqn:pauli_operators}
\end{eqnarray}
Also recall from the main text that the MZM-parity operator is
\begin{eqnarray}
    \hat{P}_t = - \hat{\gamma}_{1,t} \hat{\gamma}_{2,t} \hat{\gamma}_{3,t} \hat{\gamma}_{4,t}.
    \label{eqn:parity_op_rep}
\end{eqnarray}
Then the qubit is defined entirely by the expectation values of the Pauli operators within the even-MZM-parity subspace $\langle\hat{P}_t\rangle = +1$ \cite{Alase_2024_Phys_Rev_Res_6}. I.e. we define a qubit $``0"$ subspace as satisfying  $\langle \hat{Z} \rangle = +1,~  \langle \hat{P}_t \rangle = +1$ and a qubit $``1"$ subspace satisfying $\langle \hat{Z} \rangle = -1,~  \langle \hat{P}_t \rangle = +1$. These subspaces respectively include the ground states $\ket{0_t}$ and $\ket{1_t}$, however they also include excited states with quasiparticles, which we consider as having no impact on the qubit state (prior to any possible poisoning events). 

Given this definition of the qubit, the leakage into the odd-MZM-parity subspace $L_{\text{odd}}$ (this is equivalently the leakage into states with odd numbers of quasiparticles since the initial MZM parity is even, as defined in Eq. 6 in the main text) is in fact the qubit leakage in the conventional sense. Whereas the leakage into the even-MZM-parity excited states $L_{\text{even}}(t)$ (or equivalently the leakage into states with an even number of quasiparticles, as defined in Eq. 9 in the main text) does not coincide with qubit leakage, but is instead a mechanism which may give rise to Pauli errors, as discussed in the main text. However, if one naively defined the computational subspace exclusively as $\text{span}\{\ket{0}_t, \ket{1}_t\}$ then the appropriate qubit leakage quantity is $L_g(t) = L_{\text{odd}}(t) + L_{\text{even}}(t)$. 

We also remark that we have defined the Pauli operators, the MZM-parity operator, and in turn the computational subspace in terms of the instantaneous eigenstates of $\hat{H}$. Another, potentially appropriate computational basis may be defined in terms of the Pauli operators and MZM-parity operator at the time of qubit initialisation, i.e. at time $t = 0$. Whether the \textit{instantaneous} or \textit{initial} computational basis is most appropriate depends strictly on experimental constraints and measurement techniques. Note that the instantaneous basis will have significantly less leakage in the near-adiabatic regime, since in the true adiabatic limit a state prepared in the initial computational basis at $t = 0$ will adiabatically evolve into the equivalent state in the instantaneous basis at time $t$. Motivated by this, in the main text we made the potentially optimistic assumption that the instantaneous basis can be used in practice, and in turn we have employed the instantaneous computational basis definition.

\subsection{Further Kitaev Chain Details}
Here we detail the dependence of the band-gap in the Kitaev chain on the chemical potential $\mu$ and the analytical form of the bulk quasiparticles above the band-gap. 
In this subsection we only treat a single time-independent Kitaev chain and so we drop the $t$-dependence and $\lambda$ index for brevity. 

The bulk spectrum of the Kitaev chain may be solved by placing periodic boundary conditions on the Kitaev chain Hamiltonian (Eq. \ref{eqn:kitaev_chain_repro}), by connecting the chain in a loop and removing the ends. It is then convenient to solve the Hamiltonian in momentum space by writing the Kitaev Chain Hamiltonian as
\begin{eqnarray}
    \hat{H}_{KC} = \frac{1}{2} \mathbf{\hat{c}_k}^\dagger \mathcal{H}_{KC}(k) \mathbf{\hat{c}_k},
    ~~~ \mathbf{\hat{c}_k} = \left(\hat{c}_k^\dagger, \hat{c}_{-k} \right)
    \label{eqn:KC_momentum}
\end{eqnarray}
where $\hat{c}_k = \frac{1}{\sqrt{N}}\sum_{j} e^{-i j k} \hat{c}_j$ for $k$ in the first Brillouin zone ($k \in [-\pi, \pi]$ assuming a lattice constant of $1$) and
\begin{eqnarray}
    \mathcal{H}_{KC}(k) = 
    \begin{pmatrix}
        - \mu  -2 w\, \text{cos}( k)   &   i 2 \Delta \, \text{sin}( k )\\
        -i 2\Delta \, \text{sin}( k) &       \mu + 2 w \, \text{cos}( k )
    \end{pmatrix}.
    \label{eqn:KC_BdG_momentum}
\end{eqnarray}

Diagonalizing ${H}_{KC}(k)$ immediately gives the bulk excitation energies,
\begin{eqnarray}
\label{eq:ebulk}
    E_{\text{bulk}}(k) = \sqrt{ \left[\mu + 2 w \, \text{cos}( k ) \right] ^2 + 
    4 \Delta^2 \, \text{sin}^2( k ) }.
\end{eqnarray}
The band-gap to the bulk quasiparticles in the long chain limit is given by the minimum of $E_{\text{bulk}}(k)$ where $k$ is treated as a continuous variable. For $w = \Delta$ (as considered in the main text), the minimum occurs at $k = 0$, therefore the band-gap is
\begin{eqnarray}
    E_{\text{gap},\, \Delta = w} =  |2w - \mu|.
\end{eqnarray}
In Fig. \ref{fig:KC_excitation_spectrum} we show the bulk gap in the infinite chain length limit along with the excitation energies for a finite chain of size $N = 40$ (as considered in Fig. 2 in the main text). At the topological phase transition ($\mu = 2|w|$) the band-gap closes and at $|\mu| > 2|w|$ the Kitaev chain is in the trivial phase. Furthermore, the infinite length band-gap equation is a reasonable approximation to the $N = 40$ band-gap for $\mu \lesssim 0.9 \times 2 |w|$, which applies to all scenarios considered in the main text. Note that the band-gap for the finite chain is therefore approximately linear in $\mu$. Therefore in the chemical potential ramp scenario considered in the main text, since $\mu(t) = v t$ and $w = \Delta > 0$ (where $v$ is the chemical potential ramp), we also have that $ E_{\text{gap}} \approx 2 w - v t$.

\begin{figure}[h!]
\includegraphics[width=6cm]{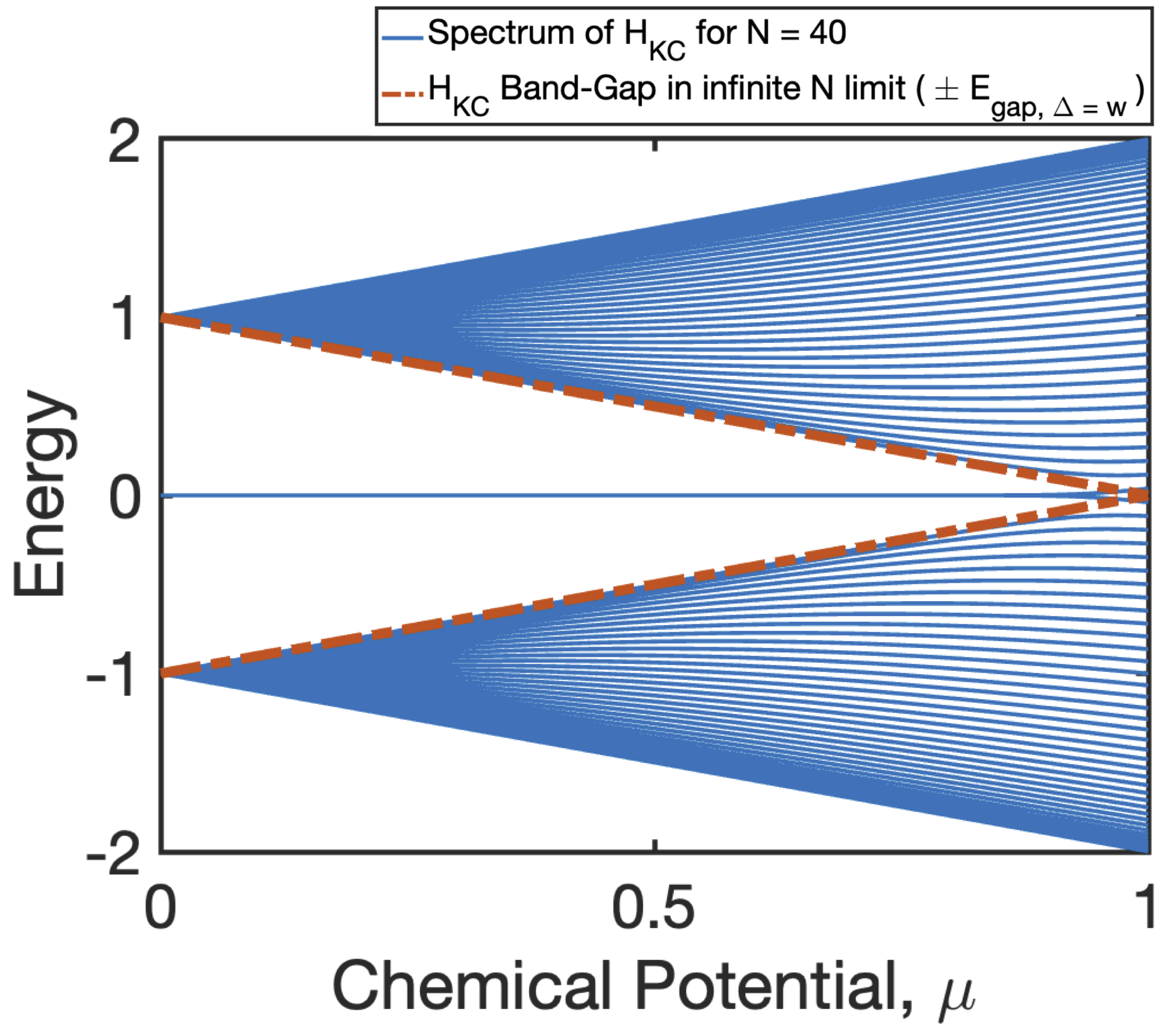}
\caption{Excitation Spectrum for the Kitaev Chain, i.e. the eigenvalues of the Bogoliubov-de Gennes matrix $H_{\text{KC}}$, as a function of $\mu$ (blue lines) and the band-gap in the infinite chain length limit (dashed orange) given by $E_{\text{gap},\, \Delta = w}$. As discussed in Appendix~\ref{sec:app_a}, the spectrum of $H_{\text{KC}}$ is symmetric around zero-energy $E = 0$. The excitation energies of the bulk quasiparticles, and the near-zero fermionic mode, are given by the positive energy eigenvalues. Since this plot shows both the positive and negative eigenvalues, $\pm E_{\text{gap},\, \Delta = w}$ (also in orange) is also shown for completeness.} 
\label{fig:KC_excitation_spectrum}
\end{figure}

\subsection{The Wannier Quasiparticles} \label{sec:wannierqp}
In this section, we recall the definition and some properties of Wannier quasiparticles (QPs) 
that we later use for proving some results on leakage. Consider a single Kitaev chain
in the topological regime described by the Hamiltonian $\hat{H}_{\rm KC}$.
Let $\{\hat{d}_k, \hat{d}^\dagger_k\}$ for $k=0,\dots,N-1$ denote the QP annhiliation and
creation operators, with $k=0$ corresponding to the zero-energy mode by convention.
Then Wannier quasiparticles under open boundary conditions are described by annihilation and 
creation operators $\{\hat{\eta}_j, \hat{\eta}^\dagger_j\}$ for $j=1,\dots,N-1$, satisfying the 
following conditions:
\begin{enumerate}
    \item Each $\hat{\eta}_j$ is a linear combination
    of $\{\hat{d}_k\}$ for $k=1,\dots,N-1$, i.e. annihilation
    operators corresponding to the bulk (above-gap) quasiparticles.
    Consequently, each $\hat{\eta}_j^\dagger$ is a linear combination
    of $\{\hat{d}_k^\dagger\}$ for $k=1,\dots,N-1$.
    \item The operators $\{\hat{\eta}_j, \hat{\eta}_j^\dagger\}$ satisfy canonical anticommutation rules.
    \item The operators $\{\hat{\eta}_j\}$ span the space $\text{span}\{\hat{d}_k\}$. Equivalently,
    \begin{equation}
        \sum_{j=1}^{N-1}\hat{\eta}_j^\dagger \hat{\eta}_j = \sum_{k=1}^{N-1}\hat{d}_k^\dagger \hat{d}_k.
    \end{equation}
    \item Each operator $\hat{\eta}_j$ is localized around site $j$ of the chain. 
\end{enumerate}

In Ref.~\cite{Alase_2024_Phys_Rev_Res_6}, it was shown that $\hat{\eta}_j$ can be chosen 
to be exponentially localized around a position $x_j$, which typically lies between
sites $j$ and $j+1$ for clean systems. 
This is a general result that relies only on the presence of
a topological band gap.

Recall that by definition $\ket{\Omega}$ (in this subsection we drop the $t$ index for convenience) is the vacuum of quasiparticles $\hat{d}_k$. Therefore the two nearly degenerate ground states $\ket{\Omega}$ 
and $\hat{d}_0^\dagger \ket{\Omega}$ of $\hat{H}_{\rm KC}$ satisfy
\begin{equation}
    \hat{d}_k\ket{\Omega} = \hat{d}_k(\hat{d}_0^\dagger\ket{\Omega}) = 0,\quad
    k = 1,\dots,N-1.
\end{equation}
Then by linearity, the two ground states also satisfy
\begin{equation}
    \hat{\eta}_j\ket{\Omega} = \hat{\eta}_j(\hat{d}_0^\dagger\ket{\Omega}) = 0,\quad
    k = 1,\dots,N-1.
\end{equation}

\subsection{Definition of $L_{\rm even}$ and $L_{\rm odd}$}

Let $\ket{\Psi(0)}$ be the initial state of the tetron. 
We assume that the tetron is initialized in a ground state with 
total even MZM-parity, i.e., 
\begin{equation}
\label{eq:initialstate}
    \hat{H}(0)\ket{\Psi(0)} = E_{0,t=0}\ket{\Psi(0)} + O(e^{-N/\xi})
\end{equation}
and
\begin{equation}
\hat{P}_0\ket{\Psi(0)} = \ket{\Psi(0)},
\end{equation}
where 
Eq.~(\ref{eq:initialstate}) includes an energy splitting that is exponentially small in the ratio of the Kitaev chain length ($N$) and the superconducting coherence length ($\xi$), $E_{0,t=0}$ is the ground state energy at $t=0$, and
$\hat{P}_0 = (i\hat{\gamma}_{1,t=0}\hat{\gamma}_{2,t=0})(i\hat{\gamma}_{3,t=0}\hat{\gamma}_{4,t=0})$.

Recall that $\hat{H}(t) = \sum_{\lambda =1 }^2 \sum_{i = 1}^{N}  \varepsilon_{k,t}^{(\lambda)} \hat{d}_{k,t}^{(\lambda)\dagger} \hat{d}_{k,t}^{(\lambda)} + E_{0,t}$, where $\varepsilon_0^{(1)} \approx \varepsilon_0^{(2)} \approx 0$. 

Let $\hat{\nu}_t = \sum_{\lambda=1}^2\sum_{k=2}^{N-1}\hat{d}^\dagger_{k,t} \hat{d}_{k,t}$ denote the bulk quasiparticle number operator.
Let $\hat{P}_{\nu, p,t}$ denote the projector on the sector of the Fock space
with $\nu$ excited bulk quasiparticles and parity $p \in \{1,-1\}$
for the MZM parity operator, for quasiparticles and MZMs of $\hat{H}(t)$. In other words, $\hat{P}_{\nu,p,t}\ket{\Psi(t)} = \ket{\Psi(t)}$ 
if 
\begin{equation}
    \hat{\nu}_t\ket{\Psi(t)} = \nu \ket{\Psi(t)} \quad \text{and} \quad
    \hat{P}_t\ket{\Psi(t)} = p\ket{\Psi}(t).
\end{equation}
A non-zero value of $\nu$ indicates leakage into excited states,
whereas $p=-1$ reflects change in the joint parity of the MZMs.
We now define 
\begin{equation}
    L_{\nu,p} = \expval{\Psi(t)|\hat{P}_{\nu,p}|\Psi(t)}
\end{equation}
as the leakage quantities of interest. As previously discussed, from the point of view of error correction, the leakage from the tetron
qubit is 
\begin{equation}
    L_{\rm odd} = \sum_{\nu=0}^{2N-2} L_{\nu,-1}.
\end{equation}
However, we focus on the leakage out of the ground subspace, which is
\begin{equation}
    L_g = \sum_{\nu=1}^{2N-2} L_{\nu,1} + L_{\nu,-1}.
\end{equation}
This is because our analysis does not take into account the thermal processes
that allow absorption of bulk quasiparticles by MZMs. In the experimentally
relevant parameter regimes, nearly all excited quasiparticles are absorbed by
MZMs, which can lead to Pauli errors~\cite{karzig_2021_phys_rev_let_126}. Therefore it is important to 
characterize $L_g$.

In this work, we only focus on total parity-preserving error processes.
Assuming a perfect initialization, the initial state resides in the
$\nu=0,~p=1$ sector. The evolved state $\ket{\Psi(t)}$ then must satisfy 
\begin{equation}
    (-1)^{\hat{\nu}}\hat{P}_t = 1.
\end{equation}
Therefore we have $L_{\nu,-(-1)^\nu}=0$. Consequently, we only need to analyze
the quantities $L_{\nu,(-1)^\nu}$, which we now rename as $L_\nu := L_{\nu,(-1)^\nu}$.
However, we remark that this nomenclature is not suitable for the analysis 
of total parity-violating processes, such as extrinsic QPP.

Consequently, we now have
\begin{equation}
    L_{\rm odd} = \sum_{\nu \, {\rm odd}} L_{\nu},
\end{equation}
and we can similarly define 
\begin{equation}
    L_{\rm even} = \sum_{\nu\, {\rm even}, \nu\ne 0} L_{\nu}.
\end{equation}
Moreover, we can now express $L_g$ as 
\begin{equation}
    L_g = L_{\rm even} + L_{\rm odd}.
\end{equation}
Since the time-dependent Hamiltonian is quadratic in annihilation and creation operators,
typically leakage processes with fewer quasiparticles dominate in the low-leakage regime. 
We therefore have
\begin{equation}
    L_{\rm odd} \approx L_1,\quad L_{\rm even} \approx L_2.
\end{equation}

\section{Solving the Kitaev-Tetron using the Covariance Matrix Formalism}
\label{sec:app_b}
As discussed in the main text, we use the covariance matrix formalism \cite{Surace_2022_SciPost_Lec, Mishmash_2020_PRB_101} to efficiently numerically solve the time evolution of the leakages into the sectors with even ($L_{\text{even}}$) and odd  ($L_{\text(odd)}$) numbers of quasiparticles (QPs), in the Kitaev-tetron qubit. For a pedagogical introduction to this method, please see Ref.\ \cite{Surace_2022_SciPost_Lec}. Here, we present the relevant elements of this formalism to our calculation. Note that our notation and equations differ slightly in form from Ref.\ \cite{Surace_2022_SciPost_Lec} as our notation explicitly applies to two 1D lattices and we use an alternate ordering of operators in Eq.~(\ref{eqn:c_vec}) below.

We consider the time evolution of the Kitaev-tetron $\hat{H}(t)$ defined on $2N$ lattice sites. The full Hilbert space of this problem is the $2^{2N}$ dimension Fock space $\mathcal{F}_{2N}$. However, in the covariance matrix method we solve the time evolution of a $4N \times 4N$ covariance matrix or equivalently the $4N \times 4N$ correlation matrix (these matrices are related in a simple manner) in the $4N$ dimension Bogoliubov-de Gennes space $\mathcal{H}_{\text{BdG}}$ (as defined in Appendix~\ref{sec:app_a}). In our simulation we solve the time-evolution of the correlation matrix, the elements of which are
\begin{eqnarray}
     &\Gamma_{i,j}^{c^\dagger c (\lambda, \zeta)}(t) =
     \langle \Psi(t) |\hat{c}_i^{(\lambda)\dagger} \hat{c}_j^{(\zeta)}| \Psi(t)\rangle, \nonumber \\
     &\Gamma_{i,j}^{c^\dagger c^\dagger(\lambda, \zeta)}(t) = 
     \langle \Psi(t) |\hat{c}_i^{(\lambda)\dagger} \hat{c}_j^{(\zeta) \dagger}| \Psi(t)\rangle, \nonumber \\   
      &\Gamma_{i,j}^{c c (\lambda, \zeta)}(t) = 
      \langle \Psi(t) |\hat{c}_i^{(\lambda)} \hat{c}_j^{(\zeta)}| \Psi(t)\rangle, \nonumber \\
      &\Gamma_{i,j}^{c c^\dagger (\lambda, \zeta)}(t) = 
      \langle \Psi(t) |\hat{c}_i^{(\lambda)} \hat{c}_j^{(\zeta)\dagger}| \Psi(t)\rangle, \nonumber 
\end{eqnarray}
where $\ket{\Psi(t)}$ is the state of the tetron at time $t$, $\hat{c}_i^{(\lambda)}$ is the fermionic annihilation operator on site $i$ on chain $\lambda \in \{1, 2\}$, and $\hat{c}_i^{(\lambda) \dagger}$ is the fermionic creation operator on site $i$ on chain $\lambda$. Note that these operators satisfy the usual fermionic anticommutation relations $\{\hat{c}_i^{(\lambda)}, \hat{c}_j^{(\zeta) \dagger} \} = \delta_{i,j} \delta_{\lambda,\zeta}$ and $\{\hat{c}_i^{(\lambda)}, \hat{c}_j^{(\zeta)} \} = 0$, where $\delta_{i,j}$ is the kronecker-delta and $\{A, B\} = AB + BA$. We construct the full $2N \times 2N$ block matrices describing correlations between chains $\lambda$ and $\zeta$ as
\begin{eqnarray}
    \Gamma^{(\lambda, \zeta)}(t) = 
    \left( \begin{matrix}
        \Gamma^{c^\dagger c (\lambda, \zeta)}(t)    & \Gamma^{c^\dagger c^\dagger (\lambda, \zeta)}(t) \\
        \Gamma^{c c (\lambda, \zeta)}(t) & \Gamma^{c c^\dagger (\lambda, \zeta)}(t)
    \end{matrix} \right),
\end{eqnarray}
and we construct the full $4N \times 4N$ correlation matrix (in the site-local basis) as
\begin{eqnarray}
    \Gamma(t) = 
    \left( \begin{matrix}
        \Gamma^{(1,1)}(t)  &   \Gamma^{(1,2)}(t) \\
        \Gamma^{(2,1)}(t)  &   \Gamma^{(2,2)}(t)
    \end{matrix} \right).    
\end{eqnarray}

As discussed in the main text, we initialize the qubit in the $\ket{\Psi(0)} = \ket{+} = \frac{1}{\sqrt{2}}\left(\ket{0_0} + \ket{1_0}\right)$ state (where $\ket{0_t}$ and $\ket{1_t}$ are the instantaneous ground states with even-even and odd-odd parity, respectively, on each Kitaev chain). We construct the corresponding correlation matrices for the $\ket{0_0}$ and $\ket{1_0}$ states in the instantaneous Bogololibov quasiparticle basis, we call these $\Upsilon_{\ket{0}}$ and $\Upsilon_{\ket{1}}$ (in the instantaneous basis these matrices and independent of $t$). These are the matrices defined equivalently to above with the site fermionic creation and annihilation operators replaced with the Bogoliubov quasiparticle operators $\hat{c}_i^{(\lambda)} \to \hat{d}_i^{(\lambda)}$ and $\hat{c}_i^{(\lambda)\dagger} \to \hat{d}_i^{(\lambda)\dagger}$, where the $\hat{d}_i^{(\lambda)}$ and $\hat{d}_i^{(\lambda)\dagger}$ are as defined in Eq. (2) of the main text and in Appendix~\ref{sec:app_a}. 

The $2N \times 2N$ block matrices for $\Upsilon_{\ket{0}}$ are $\Upsilon_{\ket{0}}^{(1,2)} = \Upsilon_{\ket{0}}^{(2,1)} = \mathbf{0}$ and 
\begin{eqnarray}
     \Upsilon_{\ket{0}}^{(1,1)} = \Upsilon_{\ket{0}}^{(2,2)} = 
    \left( \begin{matrix} 
        0_{N}   & 0_{N} \\
        0_{N}   & \mathds{1}_N 
    \end{matrix} \right).   
\end{eqnarray}
Whereas the block matrices for $\Upsilon_{\ket{1}}$ are $\Upsilon_{\ket{1}}^{(1,2)} = \Upsilon_{\ket{1}}^{(2,1)} = \mathbf{0}$ and 
\begin{eqnarray}
     \Upsilon_{\ket{1}}^{(1,1)} = \Upsilon_{\ket{1}}^{(2,2)} = 
    \left( \begin{matrix} 
        \text{diag}\left(0, 0, ..., 0, 1 \right)   & 0_{N} \\
        0_{N}   & \text{diag}\left(0, 1, 1, ..., 1, 1 \right)  
    \end{matrix} \right).   
\end{eqnarray}
The correlation matrix for the initial state $\ket{\Psi(0)} = \ket{+}$ in the instantaneous Bogololibov quasiparticle basis is then
\begin{eqnarray}
    \Upsilon_{\ket{+}} = \frac{1}{2} \left(\Upsilon_{\ket{0}} + \Upsilon_{\ket{1}} + \Upsilon_{\ket{+}}^{(\text{off-diag})} + \Upsilon_{\ket{+}}^{(\text{off-diag})\dagger} \right),
\end{eqnarray}
Where $\Upsilon_{\ket{+}}^{(\text{off-diag})}$ is a $4N \times 4N$ matrix, with elements
\begin{eqnarray}
    \Upsilon_{\ket{+}j,k}^{(\text{off-diag})} = 
     \begin{cases}
       +i, & j =N, k = 2N+1   \\
       -i, & j =2N, k = N+1 \\
       0,  & \text{otherwise.}
     \end{cases}
\end{eqnarray}
The basis of the correlation matrix can be conveniently rotated using the normalized eigenvectors of the Bogoliubov-de Gennes (BdG) Hamiltonian $H(t)$ (defined in Eq. \ref{eqn:bdg_form}) corresponding to the tetron Hamiltonian $\hat{H}(t)$. Let $V_t$ be a $4N \times 4N$ size unitary matrix columns with columns that are the normalised eigenvectors of $H(t)$, namely $\{|d_{k, t}^{(\lambda)}\rangle, \tau_x \kappa |d_{k, t}^{(\lambda)}\rangle \}$ for $k = 0, \cdots, N-1$ and $\lambda \in \{1,2\}$ (as introduced in Appendix~\ref{sec:app_a}). Then, 

\begin{eqnarray}
   \Gamma(t) = V_t^{\text{*}} \Upsilon(t) V_t^\text{T},
\end{eqnarray}
where $^{\text{*}}$ denotes the complex conjugate and $^\text{T}$ denotes the matrix transpose. The time evolution of the correlation matrix is given by, 
\begin{eqnarray}
    \Gamma(t) =  U(t, 0) \, \Gamma(0)\, U(t,0)^\dagger,
\end{eqnarray}
where
\begin{eqnarray}
    U(t,0) = \mathcal{T} \left[ e^{i \int_{0}^{t} d \tau H(\tau)} \right]
\end{eqnarray}
and $\mathcal{T}$ is the time ordering operator. This can be numerically approximated by calculating $\Gamma(t)$ at discrete time-steps $t_i \in [0, \delta t, 2 \delta t, ..., (N_{\text{steps}}-1)\delta t]$, where $\delta t$ is the time-step size and $N_{\text{steps}}$ is the number of time-steps. The correlation matrix can be numerically approximated between successive time-steps using the BdG matrix as
\begin{eqnarray}
    \Gamma(t_i + \delta t) \approx e^{i H(t_i) \delta t} \Gamma(t_i) e^{- i H(t_i) \delta t}.
\end{eqnarray}
To calculate the leakage quantities discussed in the main text, we make use of the covariance matrix, which is closely related to the correlation matrix. It is convenient to define the covariance matrix in terms of Majorana operators. For our numerical calculations, we use rescaled Majorana operators (which correspond to normalized vectors in the BdG space) which differ from the Majorana operators defined in the text by a factor of $1/\sqrt{2}$. The rescaled Majorana operators defined at each lattice site are,
\begin{eqnarray}
    \hat{r}_i^{(\lambda)} = \frac{\hat{c}_i^{(\lambda)} + \hat{c}_i^{(\lambda) \dagger}}{\sqrt{2}}, \hspace{0.5cm}
    \hat{r}_{i+N}^{(\lambda)} = \frac{\hat{c}_i^{(\lambda)} - \hat{c}_i^{(\lambda) \dagger}}{i \sqrt{2}},
\end{eqnarray}
where $i \in \{1, 2, ..., N\}$ and these rescaled Majorana operators satisfy the rescaled Majorana anticommutation relation $\{\hat{r}_i, \hat{r}_j\} = \delta_{i,j}$. We collect all rescaled Majorana operators in a single vector
\begin{eqnarray}
    \mathbf{\hat{r}} = \left(\hat{r}_1^{(1)}, \hat{r}_2^{(1)}, ..., \hat{r}_{2N}^{(1)}, \hat{r}_1^{(2)}, \hat{r}_2^{(2)}, ..., \hat{r}_{2N}^{(2)} \right)^{T} = \Omega \mathbf{\hat{c}}.
\end{eqnarray}
where,
\begin{eqnarray}
        \Omega = \mathds{1}_2 \otimes \frac{1}{\sqrt{2}} \left( \begin{matrix}
        \mathds{1}_N     & \mathds{1}_N \\
        -i \mathds{1}_N  & i\mathds{1}_N
    \end{matrix} \right),
\end{eqnarray}
and $\mathbf{\hat{c}}$ is defined in Eq.~(\ref{eqn:c_vec}) in Appendix~\ref{sec:app_a}. The $4N \times 4N$ covariance matrix is given by
\begin{eqnarray}
    M_{i,j}(t) = -i\, \bra{\Psi(t)} [\hat{r}_i, \hat{r}_j] \ket{\Psi(t)},
\end{eqnarray}
where $[A,B] = AB - BA$ denotes the commutator. The covariance matrix is related to the correlation matrix by,
\begin{eqnarray}
    &M(t)
    = - i \, \Omega^{\text{*}}\left( 2 \Gamma(t) -  \mathds{1}_{4N} \right)\Omega^{\text{T}}.
    \label{eqn:cov_from_corr}
\end{eqnarray}

Using Wick's theorem, expectation values of any even product of fermionic operators can be extracted from the correlation and covariance matrix \cite{bravyi_2012_comMathPhys_316, Surace_2022_SciPost_Lec}. Specifically, let $q_1, ..., q_{4N}$ be rescaled Majorana operators in an arbitrary basis (these are appropriate superpositions of the $\hat{r}_j^{(\lambda)}$ which satisfy the rescaled Majorana anticommutation rules and form a complete basis). Then the expectation value of an even product of $2n$ of these operators is
\begin{eqnarray}
    \expval{(-2i)^n q_1 q_2 \cdots q_{2n}} = \text{Pf}(M'|_{q_1q_2 \cdots q_{2n}}),
\end{eqnarray}
where $\text{Pf}$ denotes the Pfaffian and $M'|_{q_1q_2 \cdots q_{2n}}$ is the covariance matrix in the basis of $q_1, ..., q_{4N}$ restricted to the rows and columns corresponding to the operators $q_1, q_2, \cdots q_{2n}$. This lets us calculate the expectation of the MZM-parity $\hat{P}_t$. To do this we introduce the rescaled Majorana operators of the instantaneous quasiparticle basis as
\begin{eqnarray}
    \hat{p}_{i,t}^{(\lambda)} = \frac{\hat{d}_{i,t}^{(\lambda)} + \hat{d}_{i,t}^{(\lambda) \dagger}}{\sqrt{2}}, \hspace{0.5cm}
    \hat{p}_{i+N,t}^{(\lambda)} = \frac{\hat{d}_{i,t}^{(\lambda)} - \hat{d}_{i,t}^{(\lambda) \dagger}}{i \sqrt{2}}.
    \label{eqn:Majorana_qp}
\end{eqnarray}
The covariance matrix expressed in this basis is then simply given by Eq.~(\ref{eqn:cov_from_corr}) with the substitution $\Gamma(t) \to \Upsilon(t)$. The expectation value of the MZM-parity $P_t = - \hat{\gamma}_{1,t} \hat{\gamma}_{2,t} \hat{\gamma}_{3,t} \hat{\gamma}_{4,t} = -4\hat{p}_{1,t}^{(1)}\hat{p}_{N+1,t}^{(1)}\hat{p}_{1,t}^{(2)}\hat{p}_{N+1,t}^{(2)}$ is then given by 
\begin{eqnarray}
    \bra{\Psi(t)}\hat{P}_t \ket{\Psi(t)} = \bra{\Psi(t)}-4\hat{p}_{1,t}^{(1)}\hat{p}_{N+1,t}^{(1)}\hat{p}_{1,t}^{(2)}\hat{p}_{N+1,t}^{(2)} \ket{\Psi(t)} =
    \text{Pf}\left(\Xi(t)|_{\hat{p}_{1,t}^{(1)}\hat{p}_{N+1,t}^{(1)}\hat{p}_{1,t}^{(2)}\hat{p}_{N+1,t}^{(2)}}\right).
\end{eqnarray}
where $\Xi(t)$ is the covariance matrix expressed in the instantaneous quasiparticle basis. This in turn allows us to calculate the leakage into the odd-QP number sector, $L_\text{odd}(t)$, as given by Eq. (6) in the main text. To calculate the ground-subspace leakage $L_g(t)$ as defined in Eq. (7) in the main text, we use the following expression for calculating overlaps of two quantum states $\ket{\Psi_A}$ and $\ket{\Psi_B}$ from their respective covariance matrices $M_A$ and $M_B$ \cite{bravyi_2012_comMathPhys_316}
\begin{align}
   |\bra{\Psi_A} \ket{\Psi_B}|^2 = \frac{1}{2^{2N}} \sqrt{ \text{det}\left(M_A + M_B \right)}. 
\end{align}
Then the leakage into the even-QP number sector $L_{\text{even}}(t)$ is simply given by $L_{\text{even}(t)} = L_g(t) - L_{\text{odd}}$.

We remark that when diagonalizing the BdG Hamiltonian $H$ to obtain the $\hat{d}_{j,t}^{(\lambda)}$ operators and in turn the $\hat{p}_{j,t}^{(\lambda)}$ operators: there exists a near-degeneracy of the eigenvectors corresponding to $\hat{d}_{1,t}^{(1)},\, \hat{d}_{1,t}^{(1)\dagger}, \hat{d}_{1,t}^{(2)},\, \hat{d}_{1,t}^{(2)\dagger}$. It is necessary to obtain the four eigenvectors which after applying Eq.~(\ref{eqn:Majorana_qp}) give us the Majorana Zero Modes (MZMs) localized on the respective boundaries of the two Kitaev chains. To achieve this we diagonalize the two Kitaev chains separately ($H$ consists of two uncoupled $2N \times 2N$ block matrices). For each chain, we then take the two lowest energy eigenvectors (one with eigenvalue $\varepsilon_{1,t}^{(\lambda)} \geq 0$ and the other with eigenvalue $-\varepsilon_{i,t}^{(\lambda)}$ ) and form the two orthogonal superpositions of these eigenvectors to assign to $\hat{d}_{1,t}^{(1)}$ and $\hat{d}_{1,t}^{(1)\dagger}$ which (1) obey the fermionic anticommutation rules and (2) in turn give us $\hat{p}_{1,t}^{(\lambda)}$ and $\hat{p}_{N+1,t}^{(\lambda)}$ that are maximally localized on each end of the Kitaev chain. 

\section{Leakage Dynamics in the Sudden Quench Regime}
\label{sec:app_c}
In this section, we provide analytic and numeric insight into the dynamics of leakage 
in the sudden regime. The analysis in this section applies
generally to tetron qubits based on one-dimensional models of topological superconductors beyond
the Kitaev chain model, thus providing evidence for wider applicability
of the behaviors observed numerically for the Kitaev tetron model 
in special parameter regimes.
We first show that in the sudden quench limit, $L_{\rm odd}$ for large $N$ approaches a constant value which depends on the product of 
overlaps of MZM wavefunctions before and 
after the quench, as given in Eq. (12) in the main text. 
We then show that $L_g$ in this limit grows linearly
in $N$. Finally, we derive the dependence of $L_{\rm odd}$ and $L_g$ on the ramp
rate in the sudden regime.

\subsection{Scaling of leakage with system size, in the sudden quench limit}

{\bf \noindent Scaling of $L_{\rm odd}$:}
We first derive an approximate formula for leakage into the odd-MZM-parity sector
$L_{\rm odd}$ in the sudden quench limit.
As the time-dependent Hamiltonian conserves total parity, the leakage states
with odd numbers of excited quasiparticles have odd Majorana parity, i.e. 
$\expval{\hat{P}_t} = -1$. The projector on this sector is given 
by $\hat{\Pi}_t = (1-\hat{P}_t)/2$. Therefore $L_{\rm odd}$ can be expressed as
\begin{equation}
L_{\rm odd}(t) = \frac{1}{2} - \frac{\expval{\hat{P}_t}}{2},\quad
\expval{\hat{P}_t} = \bra{\Psi(t)}\hat{P}_t\ket{\Psi(t)} = 
-\expval{\Psi(t)|\hat{\gamma}_{1,t}\hat{\gamma}_{2,t}\hat{\gamma}_{3,t}\hat{\gamma}_{4,t}|\Psi(t)}.
\end{equation}
Here $\ket{\Psi(t)} = \hat{U}_t\ket{\Psi(0)}$ is the evolved state of the tetron
with $\hat{U}_t$ denoting the many-body time propagator, and 
$\{\hat{\gamma}_{l,t}\}$ for $l=1,\dots,4$ denote the instantaneous 
maximally localized MZMs at time $t$.
Then $\expval{\hat{P}_t}$ can be re-expressed as 
\begin{equation}
    \expval{\hat{P}_t} = -\expval{\Psi(0)|\hat{\gamma}_{1,t}(-t)\hat{\gamma}_{2,t}(-t)\hat{\gamma}_{3,t}(-t)\hat{\gamma}_{4,t}(-t)|\Psi(0)},
\end{equation}
where $\hat{\gamma}_{l,t}(-t) = \hat{U}_t^\dagger\hat{\gamma}_{l,t}\hat{U}_t$.
Furthermore, $\hat{\gamma}_{l,t}(-t)$ can be expanded in terms of Wannier QPs as 
\begin{eqnarray}
    \hat{\gamma}_{l,t}(-t) = \alpha_{l,0}\hat{\gamma}_{l,0}(0) + \sum_j\left( \alpha_{l,j}\hat{\eta}_{j,0}^{(\lambda)} 
    + \alpha_{l,j}^*\hat{\eta}_{j,0}^{(\lambda)\dagger}\right), ~~~~ 
    \lambda = \begin{cases} 1, & l \in \{1,2\} \\ 2, & l \in \{3,4\}
    \end{cases}
    \label{eqn:gamma_wannier_decomp}
\end{eqnarray}

Here the $t$-dependence of the coefficients $\{\alpha_{l,j}\}$ is suppressed for brevity.
If the leakage is small, we expect $0 \le 1- \alpha_{l,0} \ll 1$ and $\alpha_{l,j} \ll 1$.
Using the fact that $\hat{\eta}_{j,0}^{(\lambda)}\ket{\Psi(0)} = 0$, we obtain
\begin{eqnarray}
       \expval{\hat{P}_t} = \alpha_{1,0}\alpha_{2,0}\alpha_{3,0}\alpha_{4,0} 
    - &&\alpha_{1,0} \alpha_{2,0}
    \expval{\Psi(0)|\hat{\gamma}_{1,0}(0)\hat{\gamma}_{2,t}(0))|\Psi(t)}
    \sum_j \alpha_{3,j}\alpha_{4,j}^*
     \nonumber \\
     -
    &&\alpha_{3,0} \alpha_{4,0}
    \expval{\Psi(0)|\hat{\gamma}_{3,0}(0)\hat{\gamma}_{4,t}(0))|\Psi(t)}\sum_j \alpha_{1,j}\alpha_{2,j}^*. 
\end{eqnarray}

In the sudden limit, the time at the end of the ramp is $T = 0$ and $\ket{\hat{\eta}_{j,0}^{(\lambda)}(T)} = \ket{\hat{\eta}_{j,0}^{(\lambda)}}$.
As reviewed in Appendix~\ref{sec:app_a}, since $\ket{\hat{\eta}_{j,0}^{(\lambda)}}$ is exponentially localized around $j$,
$\alpha_{1,j}$ is exponentially small in $j$ and $\alpha_{2,j}$ is exponentially small
in $N-j$. Therefore $\alpha_{1,j}\alpha_{2,j}^*$ is exponentially small in $N$.
As a result, for sufficiently long chains, we have 
\begin{equation}
\label{eq:lpformula1}
    \lim_{N\to \infty} \expval{\hat{P}_{T}} = \alpha_{1,0}\alpha_{2,0}\alpha_{3,0}\alpha_{4,0}. 
\end{equation}
For finite $N$, $\expval{\hat{P}_{T}}$ is approximated exponentially well by
$\alpha_{1,0}\alpha_{2,0}\alpha_{3,0}\alpha_{4,0}$. 
In the vector space of linear operators, the coefficients $\{\alpha_{l,0}\}$
can be seen as overlaps
$\alpha_{l,0} = \expval{\hat{\gamma}_{l,0}|\hat{\gamma}_{l,t}(-t)} = \expval{\hat{\gamma}_{l,0}(t)|\hat{\gamma}_{l,t}(0)}$,
of Majorana wavefunctions before and after the quench. Consequently,
\begin{equation}
    \expval{\hat{P}_t} \approx \prod_l\expval{\hat{\gamma}_{l,0}(t)|\hat{\gamma}_{l,t}(0)},\quad
    L_{\rm odd} \approx \tilde{L}_{\rm odd,~sudden} \equiv (1-\prod_l\expval{\hat{\gamma}_{l,0}(t)|\hat{\gamma}_{l,t}(0)})/2
    \label{eqn:odd_sudden_approx}
\end{equation}
This formula explains the $O(1)$ behavior of $L_{\rm odd}$ 
as a function of chain length:
for sufficiently large chain lengths, $L_{\rm odd}$ is nearly independent of chain length. 
This behavior is demonstrated in the Fig.~\ref{fig:sudden_leakage_scaling} which shows the numerically computed scaling of $L_{\text{odd}}$ in $N$ for various final chemical potentials $\mu_\text{fin}$. In fact, deep in the topological phase (for $\mu_\text{fin} \ll 2|w|$) $L_{\text{odd}}$ is approximately constant in $N$ and is given by Eq.~(\ref{eqn:odd_sudden_approx}) for all chain lengths $N \geq 2$. For larger $\mu_\text{fin} = 0.5$, we find numerically that the constant scaling of $L_{\text{odd}}$ in $N$ and Eq.~(\ref{eqn:odd_sudden_approx}) hold for $N \gtrsim 10$.

\begin{figure}[h!]
\includegraphics[width=12cm]{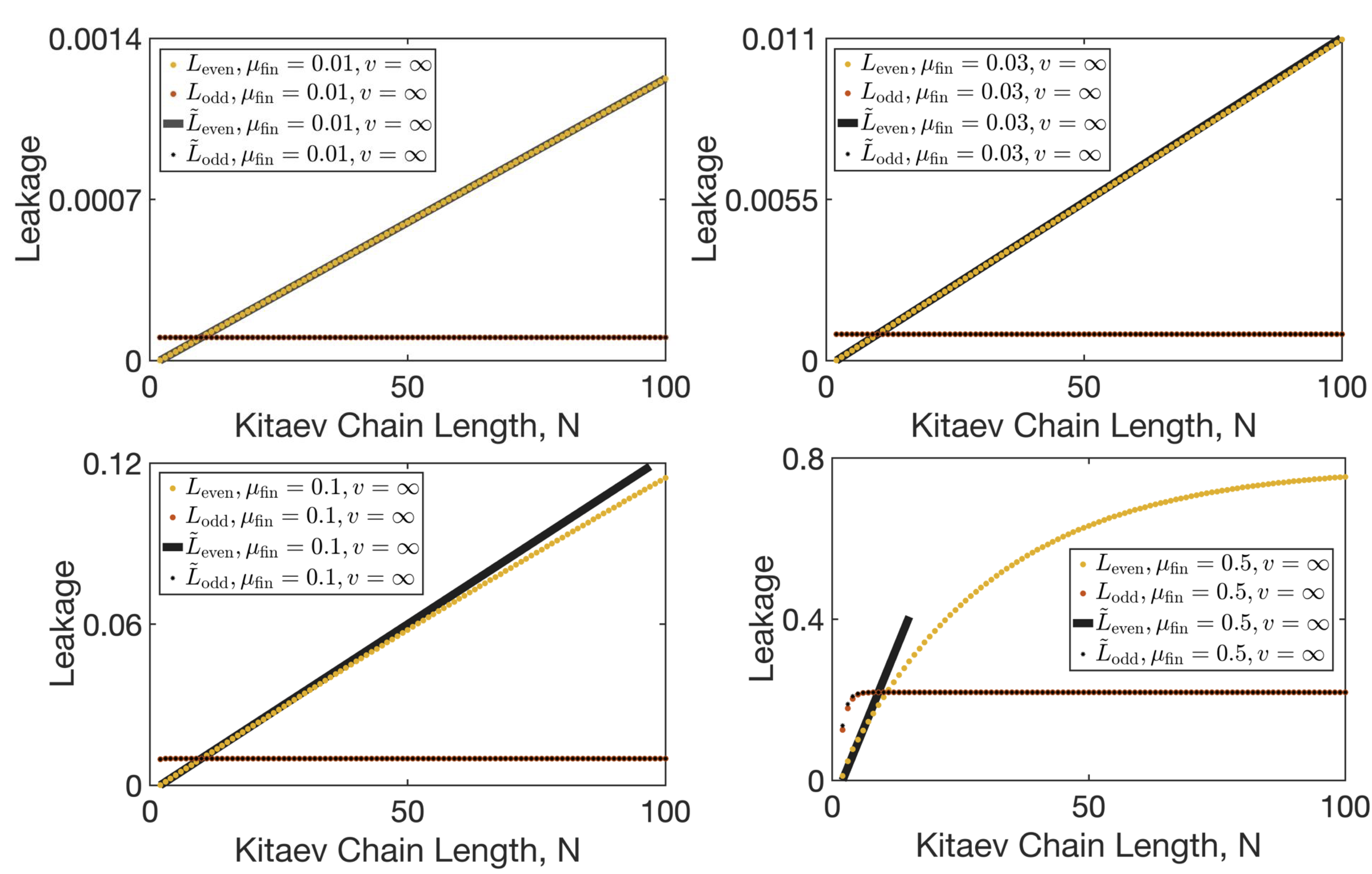}
\caption{Scaling of leakages into states with even ($L_{\text{even}}$) and odd ($L_\text{odd}$) numbers of quasiparticles in the sudden-limit, i.e. at chemical potential ramp rate $v \to \infty$ for a final chemical potential $\mu_\text{fin}$ of $0.01$ (top left), $0.03$ (top right), $0.1$ (bottom left) and $0.5$ (bottom right). All plots shown are for a Kitaev-tetron qubit with initial chemical potential $\mu_{\text{in}} = 0$, with hopping $w$ and pairing $\Delta$ of $w = \Delta = 0.5$, and chain size of $N = 40$, and with Kitaev chain lengths on the interval $N \in [2, 100]$. The leakages in the sudden-limit are computed using the covariance matrix formalism with $\mu(t)$ set to a step function at $t = 0$. The analytical approximation $\tilde{L}_{\rm even,~sudden}$ (abbreviated to $\tilde{L}_{\rm even}$ in plot legends) is given by Eq. (\ref{eqn:sudden_scaling_analytic}) and is in good agreement with $L_{\rm even}$ at $L_{\rm even} \ll 1$. The approximation $L_{\text{odd}}(v = \infty) \approx \tilde{L}_{\text{odd,~sudden}}$ in terms of the overlap of Majorana zero mode (MZMs) wavefunctions is given in Eq.~(\ref{eqn:odd_sudden_approx}) above. As shown in the panels $\tilde{L}_{\rm odd,~sudden}$ and $L_\text{odd}(v = \infty)$ are in good agreement (they are indistinguishable on these plots) with the exception of at small $N$ for when $\mu_\text{fin} = 0.5$. By definition the leakage quantities are bounded by $1$. For $\mu_\text{fin} = 0.5$ (bottom right) the tetron is no longer ``deep in the topological phase" (the scenario considered in the main text), this causes high leakages approaching $1$ and therefore $L_{\text{even}}$ saturates at significantly smaller $N$.}
\label{fig:sudden_leakage_scaling}
\end{figure}

\ 

{\bf \noindent Scaling of $L_{\rm even}$:}
To derive the scaling of $L_{\rm even}$,
it suffices to derive the scaling of $L_g = L_{\rm even} + L_{\rm odd}$.
Let us first express 
\begin{equation}
    \ket{\Psi(t)} \approx \ket{\Psi_0(t)}+\ket{\Psi_1(t)}+\ket{\Psi_2(t)},
\end{equation}
where $\ket{\Psi_q(t)}$ for $q=0,1,2,\dots$ denotes the component of the state in the sector with 
$q$ excited QPs. The components with more than $2$ QPs are assumed to have negligible norm,
as the change in chemical potential is very small.
We then have
\begin{equation}
    L_g \approx \expval{\Psi_1(t)|\Psi_1(t)} + \expval{\Psi_2(t)|\Psi_2(t)}.
\end{equation}
Note that the QP number operator satisfies,
\begin{equation}
    \nu = \expval{\Psi(t)|\hat{\nu}|\Psi(t)} = 
    \expval{\Psi_1(t)|\Psi_1(t)} + 2\expval{\Psi_2(t)|\Psi_2(t)},
\end{equation}
where $\hat{\nu} = \sum_{j,\lambda}
\hat{\eta}_{j,t}^{(\lambda)\dagger}
\hat{\eta}_{j,t}^{(\lambda)}$
is the QP number operator. Therefore, we have 
\begin{equation}
    L_g \ge \nu/2.
\end{equation}

For a small change in chemical potential and sufficiently high ramp rates, we 
can use the approximation
\begin{equation}
\label{eq:wannierqpchange}
    \hat{\eta}_{j,t}^{(\lambda)}(-t) \approx \sqrt{1-\abs{\beta_j^{(\lambda)}}^2}\hat{\eta}_{j,0}^{(\lambda)} 
    + \beta_{j}^{(\lambda)}
    \hat{\eta}_{j,0}^{(\lambda) \dagger}
    , \quad \abs{\beta_j^{(\lambda)}} \ll 1,
\end{equation}
where the $\beta$'s describe the spread of the quasiparticles.  This form is justified due to exponential localization of Wannier QPs. One can
keep contributions from Wannier operators with neighboring $j$'s to get better
approximations successively, and the following arguments can be modified accordingly
to obtain the same scaling.
Using $\hat{\eta}_{j,0}^{(\lambda)}\ket{\Psi(0)} = 0$, to first order in $\{\abs{\beta_j^{(\lambda)}}^2\}$, we get 
\begin{align}
    \nu &= \bra{\Psi(0)}\sum_{j,\lambda} \left(\sqrt{1-\abs{\beta_{j}^{(\lambda)}}^2}
    \hat{\eta}_{j,0}^{(\lambda)\dagger}
    + 
    \beta_{j}^{(\lambda)*}
    {\hat{\eta}_{j,0}^{(\lambda)}}\right)\left(\sqrt{1-\abs{\beta_{j}^{(\lambda)}}^2}\hat{\eta}_{j,0}^{(\lambda)} 
    + \beta_{j}^{(\lambda)}
    \hat{\eta}_{j,0}^{(\lambda)\dagger}
    \right) 
    \ket{\Psi(0)} \nonumber\\
    &= \sum_{j,\lambda}\abs{\beta_j^{(\lambda)}}^2.
\end{align}

For sufficiently long chains and $j$ sufficiently far away from the edges, 
$\hat{\eta}_{j,t}^{(\lambda)}(-t)$ is nearly independent of $N$. Consequently
we expect the value of $\beta_j^{(\lambda)}$ also to be nearly independent of $N$.
As the chain length increases, the number of sites $j$ in the bulk also 
increase linearly with $N$. As a result, the contribution to $\nu$, 
and consequently to $L_g$, from bulk 
sites increases linearly in $N$, as found in the numerics.
Since $L_{\rm even} = L_g - L_{\rm odd}$ and since $L_{\rm odd} \in O(1)$,
we conclude that $L_{\rm even}$ scales linearly in $N$. 

Since we have $L_{\rm odd} \approx \expval{\Psi_1(t)|\Psi_1(t)}$ nearly
independent of $N$ for large $N$, we conclude that 
$\expval{\Psi_1(t)|\Psi_1(t)} \ll \expval{\Psi_2(t)|\Psi_2(t)}$ for large $N$.
Therefore 
\begin{equation}
\label{eq:leventonqp}
    L_{\rm even} \approx \nu/2 = \sum_{j,\lambda}\abs{\beta_j^{(\lambda)}}^2/2.
\end{equation}

This behavior is demonstrated in Fig. \ref{fig:sudden_leakage_scaling}, which presents the leakage in the sudden limit for various final chemical potentials $\mu_\text{fin}$. These numerical results clearly support the conclusion that in the sudden limit $L_{\text{even}}$ scales linearly in $N$ for $L_{\text{even}}$ for sufficiently large $N$ (in fact, $N \geq 2$ appears to be sufficient) and in the low leakage limit, i.e. $L_\text{even} \ll 1$. This occurs at final chemical potentials deep in the topological phase $\mu_\text{fin} \ll 2|w|$ with intermediate chain length sizes. Importantly, it is expected that this is the desired regime for the operation of tetron qubits. However for sufficiently large $N$ or sufficiently large $\mu_\text{fin}$, the leakage $L_{\text{even}}$ will eventually saturate as $L_{\text{even}}$ approaches unity.

\

{\bf \noindent Calculation of $L_{\rm even}$ in the sudden limit for a system with periodic boundary conditions:}
Since $L_{\rm even}$ is proportional to system size, it is reasonable to calculate it by considering a system with periodic boundary conditions.
When periodic boundary conditions are applied, then the system is translationally invariant, and by Fourier transforming, the Hamiltonian can be written as a sum of noninteracting terms indexed by wavevector $k$ as in Eq.~\eqref{eqn:KC_momentum}.
Suppose one starts in the ground state of the Hamiltonian with $\mu=\mu_{in}$ and $\mu$ jumps instantaneously to the value $\mu=\mu_{\rm fin}$.
Similar to Eq.~\eqref{eq:wannierqpchange}, let us express the delocalised QP operators after the sudden quench in terms of
the ones before the quench as

\begin{equation}
    \hat{d}_{k,0^+}^{(\lambda)} \approx \sqrt{1-\abs{\beta_k^{(\lambda)}}^2} \hat{d}_{k,0}^{(\lambda)} 
    + \beta_{k}^{(\lambda)}\hat{d}_{-k,0}^{(\lambda)\dagger}, \quad \abs{\beta_k^(\lambda)} \ll 1,
\end{equation}
where $k$ belongs to the Brillouin zone.
Then from Eq.~\eqref{eq:leventonqp}, we have
\begin{equation}
    L_{\rm even} \approx 
    \frac{1}{2}
    \sum_{k, \lambda} \bra{\Psi(0)} d_{k,0^+}^{(\lambda) \dagger} d_{k,0^+}^{(\lambda)}\ket{\Psi(0)} = 
    \sum_{k,\lambda} \abs{\beta_k^{(\lambda)}}^2/2.
\end{equation}
For sufficiently large $N$, we can convert the discrete sum into integral to yield
\begin{align}
\label{eq:leventointegral}
    L_{\rm even} &\approx \frac{N}{2\pi}\sum_{k,\lambda} \left(\abs{\beta_k^{(\lambda)}}^2/2\right)(2\pi/N)\nonumber\\
    &= 
    2\frac{N}{2\pi}\int_{-\pi}^{\pi} \left(\abs{\beta_k^{(\lambda)}}^2/2\right) dk\nonumber\\
    &=
    \frac{N}{2\pi}\int_{-\pi}^{\pi} \abs{\beta_k^{(\lambda)}}^2 dk.
\end{align}

To calculate the coefficients $\{\beta_k^{(\lambda)}\}$, recall that
\begin{equation}
    H_{\rm KC}^{(\lambda)}(k) = \begin{pmatrix}
        \varepsilon_k & i\delta_k\\
        -i\delta_k & -\varepsilon_k~
    \end{pmatrix},\quad
    \varepsilon_k = -\mu-2w\cos(k), \quad \delta_k = 2\Delta\sin(k).
\end{equation}
Then the corresponding eigenvector (QP excitation operator) can be expressed as
\begin{equation}
    \hat{d}_k^{(\lambda) \dagger} = \frac{1}{\sqrt{(\varepsilon_k - E_{\rm bulk}(k))^2 + \delta_k^2}}
    \left[ -i\delta_k \hat{c}_k^\dagger + \left(\varepsilon_k  - E_{\rm bulk}(k)\right) \hat{c}_{-k} \right],
\end{equation}
with $E_{\rm bulk}(k)$ as given in Eq.~\eqref{eq:ebulk}.
For small changes in $\mu$, we obtain
\begin{equation}
    \beta_k^{(\lambda)} \approx -\frac{\delta_k(\mu_{\rm fin}-\mu_{\rm in})}{2(E_{\rm bulk}(k))^2}.
\end{equation}
Specializing to $\mu_{in}=0$, 
and for the parameter values $\Delta=w=1/2$ used in the main text,
we get
\begin{equation}
    \int_{-\pi}^{\pi} \abs{\beta_k^{(\lambda)}}^2 dk = \frac{\pi \mu_{\rm fin}^2}{4}.
\end{equation}
Substituting this value in Eq.~\eqref{eq:leventointegral} yields the leading contribution
to $L_{\rm even}$ to be
\begin{equation}
    L_{\rm even} \approx  \equiv \frac{N}{2\pi}\frac{\pi\mu_{\rm fin}^2}{4}
    = \frac{N\mu_{\rm fin}^2}{8}.
\end{equation}
To compare this against our numerical results for the tetron with open boundary conditions, we make the substitution $N \to N -2$, which gives
\begin{equation}
    L_{\rm even} \approx \tilde{L}_{\rm even,~sudden} \equiv \frac{(N - 2)}{2\pi}\frac{\pi\mu_{\rm fin}^2}{4}
    = \frac{N\mu_{\rm fin}^2}{8},
    \label{eqn:sudden_scaling_analytic}
\end{equation}
which, as demonstrated in Fig. \ref{fig:sudden_leakage_scaling}, agrees well with our numerical results for when $L_{\rm even} \ll 1$.

\
 
\subsection{Scaling of leakage with respect to ramp rate, in the sudden regime}

As stated in the main text, in our numerical study of the Kiteav-tetron qubit, we find that in the sudden regime, $L_\text{even}$ and $L_\text{odd}$ scale in $v$ according to, 
\begin{align*}
    L_{\rm odd}(v) &= L_{\rm odd}(\infty) - \frac{k_{\rm odd}}{v^2},\\
    L_{\rm even}(v) &= L_{\rm even}(\infty) - \frac{k_{\rm even}}{v^2}.
\end{align*}
Fig. \ref{fig:leak_sudden} presents this scaling for various values of $\mu_{\text{fin}}$.

\begin{figure}[h!]
\includegraphics[width=8.5cm]{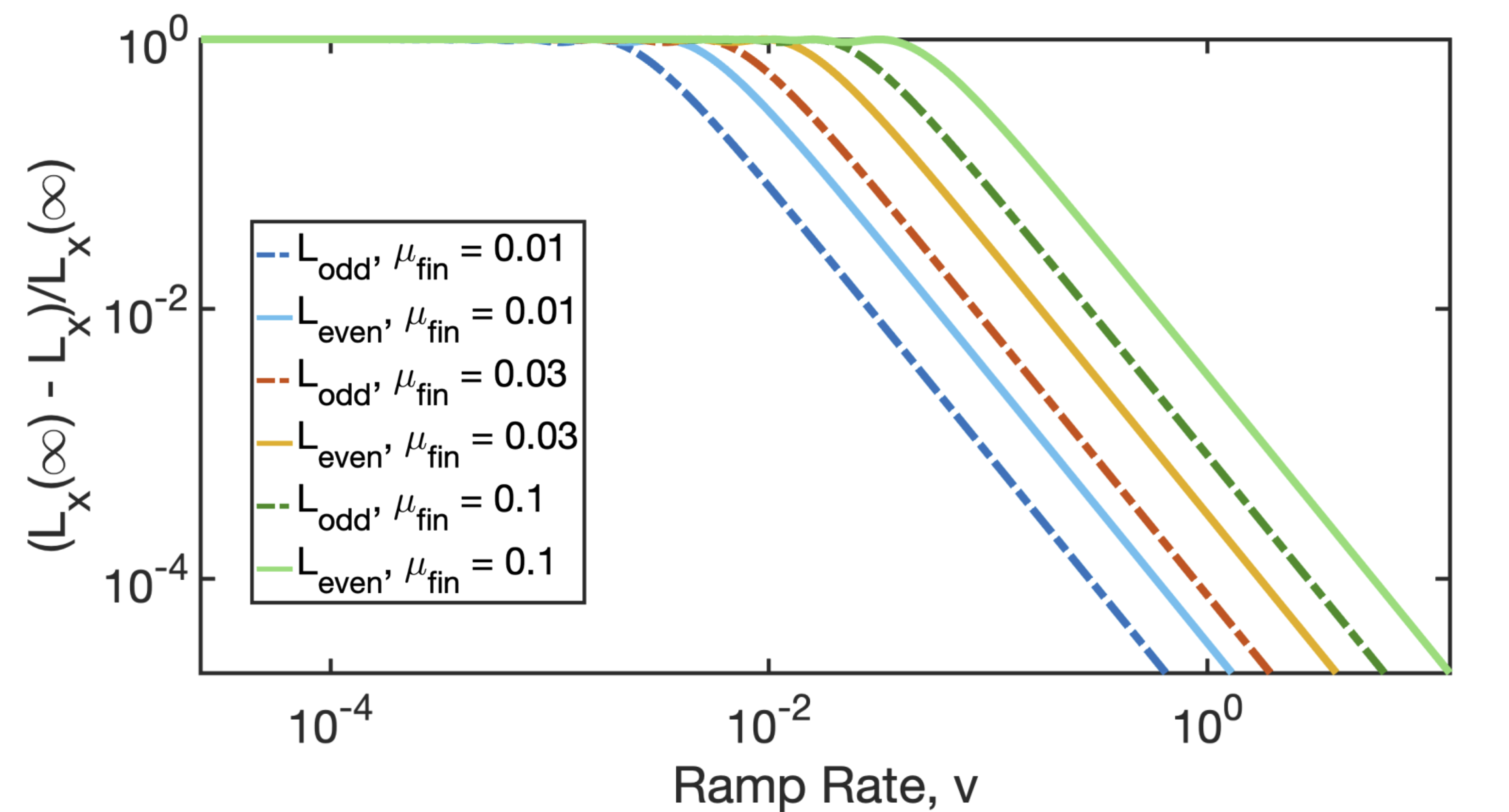}
\caption{Approach of $L_{\rm even}$ and $L_\text{odd}$ to their respective sudden-limit values $L_{\rm even}(\infty)$ and $L_{\rm odd}(\infty)$, where $x \in \{\text{even, odd}\}$, as a function 
of chemical potential ramp rate $v$: for a Kitaev-tetron qubit with initial chemical potential $\mu_{\text{in}} = 0$; final chemical potential of $\mu_\text{fin} \in \{0.01, 0.03, 0.1\}$; and hopping $w$ and superconducting pairing $\Delta$ of $w = \Delta = 0.5$. We find that the gradients in the straight line regions (on the logarithmic scale) at higher ramp rates $v$ are all $m = -2.000$ for both $L_{gc}$ and $L_{p}$.}
\label{fig:leak_sudden}
\end{figure}

\section{Leakage Dynamics in Near-Adiabatic Regime}
\label{sec:app_d}
In the main text we claim that the dynamics of the leakage quantities for a tetron undergoing a chemical potential ramp deep in the topological phase, follows half-Landau-Zener (half-LZ) physics of the form given by Eq. (4.20) and Eq. (4.30) in Ref. \cite{deGrandi_2010_Springer_75_114}. To demonstrate this, here we present numerical and analytic results consistent with half-LZ physics. 

First, according to the half-LZ physics, the tetron when prepared in a ground state (as it is in our simulations) will exhibit transition rates to excited levels which scale approximately as $v^2$ where $v$ is the chemical potential ramp rate (see Eq. 4.30 in 
 Ref.~\cite{deGrandi_2010_Springer_75_114}) with an additional oscillatory component (see Eq. 4.20 in Ref~.\cite{deGrandi_2010_Springer_75_114}).  In Fig.~\ref{fig:near_adibatic_osc} we show that both leakage quantities, namely the leakage into states with even ($L_{\text{even}}$) and odd ($L_{\text{odd}}$) numbers of quasiparticles (QPs), oscillate in $v$ around a mean-value which grows as $v^2$.

For simplicity, let us assume that the initial state is 
$\ket{\Psi(0)} = \ket{\phi_{{\bf 0},0}}$.
In the adiabatic regime marked by $v/g_{\text{bulk}}^2 \ll 1$, 
the coefficients of excited states are given by \cite{deGrandi_2010_Springer_75_114}
\begin{equation}
\label{eqn:originaladiabatic}
    \abs{a_{\bf n}(T)}^2 \approx \frac{\abs{\expval{\phi_{{\bf n},0}|\partial_{t=0}|\phi_{{\bf 0},0}}}^2}{(E_{{\bf n},0}-E_{{\bf 0},0})^2}
    + \frac{\abs{\expval{\phi_{{\bf n},T}|\partial_{t=T}|\phi_{{\bf 0},T}}}^2}{(E_{{\bf n},T}-E_{{\bf 0},T})^2}
    -2\frac{\expval{\phi_{{\bf n},0}|\partial_{t=0}|\phi_{{\bf 0},0}}}{(E_{{\bf n},0}-E_{{\bf 0},0})}
    \frac{\expval{\phi_{{\bf n},T}|\partial_{t=T}|\phi_{{\bf 0},T}}}{(E_{{\bf n},T}-E_{{\bf 0},T})}\cos(\Delta\theta_{\bf n}),
\end{equation}
where $E_{{\bf n},t}$ denotes the instantaneous energy eigenvalue of 
the state $\ket{\phi_{{\bf n},t}}$ and 
$\Delta\theta_{\bf n} = \int_{0}^{T}(E_{{\bf n},t}-E_{{\bf 0},t})dt$
is the dynamic phase difference accumulated between the state $\ket{\phi_{{\bf 0},t}}$ and
$\ket{\phi_{{\bf n},t}}$. Note that the leakage $L_{\bf n} = \abs{a_{\bf n}(T)}^2$ 
in the state $\ket{\phi_{{\bf n},T}}$ does not depend on the wavefunctions and energies
of the states $\ket{\phi_{{\bf n'},T}}$ with ${\bf n'} \ne {\bf n}$. In fact,
Eq.~\eqref{eqn:originaladiabatic} precisely governs the population transfer
in two-level half-Landau-Zener effect. Clearly, the total leakage
$L_g = \sum_{{\bf n}\ne {\bf 0}}L_{\bf n}$ equals the sum of leakages in individual
excited levels, which conform to two-level half-Landau-Zener dynamics.

To evaluate the explicit dependence of $\abs{a_{\bf n}(T)}^2$
on the ramp rate $v$, we can re-express this formula as follows.
First note that 

\begin{equation}
    \frac{\partial (\hat{H}(t) -E_{{\bf 0}, 0})}{\partial t}\ket{\phi_{{\bf 0}, 0}} + (\hat{H}(t) -E_{{\bf 0}, 0 })\frac{\partial \ket{\phi_{{\bf 0}, 0}}}{\partial 0} =0.
\end{equation}
Therefore, we have 
\begin{equation}
    \bra{\phi_{{\bf n}, 0 }}\partial_{t=0}\ket{\phi_{{\bf 0}, 0 }} = 
    \frac{\bra{\phi_{{\bf n}, 0}}(\partial \hat{H}(t)/\partial t)\ket{\phi_{{\bf0}, 0}}}{(E_{{\bf n}, 0}-E_{{\bf 0}, 0})}.
\end{equation}
As $\partial \hat{H}(t)/\partial t = v\hat{n}$, substituting this in Eq.~\eqref{eqn:originaladiabatic} immediately reveals
\begin{equation}
    \abs{a_{\bf n}(T)}^2 \propto \frac{v^2}{(E_{{\bf n}, 0}-E_{{\bf 0}, 0})^4}. 
\end{equation}

Therefore, the leakage in the adiabatic regime grows quadratically with the ramp rate
and quartically with the inverse of the bulk gap.

The final term in the right-hand-side of Eq.~\eqref{eqn:originaladiabatic} gives rise
to oscillations of $L_{\rm odd}$ and $L_g$ as a function of $v$. The oscillations
are most pronounced for small values of $\mu_{\rm fin}$. In this nearly dispersionless regime,
the energies of all quasiparticles are nearly equal and remain almost unchanged throughout
the evolution. Therefore all singly excited states contributing to $L_{\rm odd}$ acquire the same dynamical phase. 
An approximate value of this dynamical phase is given by
\begin{equation}
    \Delta\theta_{\bf n} = \int_{0}^{\mu_{\rm fin}/v} E_{\rm gap} dt 
    = \frac{E_{\rm gap}\mu_{\rm fin}}{v}.
    \label{eqn:dynamic_phase}
\end{equation}
Similarly, the doubly excited states contributing to $L_{\rm even}$ acquire twice the amount of dynamical phase due to double the energy difference. The numerical results in Fig. \ref{fig:near_adibatic_osc} for $\mu_{\text{fin}} = 0.03$ show oscillations in $L_{\text{odd}}$ in agreement with Eq.~(\ref{eqn:dynamic_phase}) and oscillations of $L_{\text{even}}$ at double the frequency. Due to the finite dispersion, these oscillations exhibit some interference and therefore do not exactly follow Eq.~(\ref{eqn:originaladiabatic}) which applies to leakage for a single level. However, as shown in panels (d) and (e) of the figure, fitting a function of the form of Eq.~(\ref{eqn:originaladiabatic}) to $L_{\text{even/odd}}$ on a restricted domain $v$ gives oscillations approximately of the form expected for a single level.

To obtain the scaling of $L_{\rm even}$ in the near-adiabatic regime, 
we use the periodic boundary conditions. Let us express the delocalised quasiparticles backwards evolved from t = T, as
\begin{equation}
    \hat{d}_{k,T}^{(\lambda)}(-T) \approx \sqrt{1-\abs{\beta_k^{(\lambda)}}^2} \hat{d}_{k,0}^{(\lambda)} 
    + \beta_{k}^{(\lambda)}\hat{d}_{-k,0}^{(\lambda)\dagger}, \quad \abs{\beta_k^{(\lambda)}} \ll 1,
\end{equation}
Then $L_{\rm even}$ is well approximated by
\begin{equation}
\label{eq:leventobeta}
    L_{\rm even} \approx \frac{N}{2\pi}\int_{-\pi}^{\pi}\abs{\beta_k^{(\lambda)}}^2 dk.
\end{equation}
Let 
\begin{equation}
    d_{k,T}^{(\lambda) \dagger}(t) = u_k^{(\lambda)}(t) c_k^{(\lambda)\dagger} + v_k^{(\lambda)}(t)c_{-k}^{(\lambda)} 
\end{equation}
and define 
\begin{equation}
    \ket{d_{k,T}^{(\lambda)\dagger}(t)} = \begin{pmatrix}u_k^{(\lambda)}(t)
    \\ v_k^{(\lambda)}(t) \end{pmatrix}.
\end{equation}
Then,
\begin{equation}
    i\hbar \frac{d \ket{d_{k,T}^{(\lambda)\dagger}(t)}}{dt} = H_k^{(\lambda)}(t)\ket{d_{k,T}^{(\lambda) \dagger}(t)}.
\end{equation}
Using Eq.~\eqref{eqn:originaladiabatic}, we get
\begin{eqnarray}
    \abs{\beta_k^{(\lambda)}}^2 \approx &&\frac{\abs{
    \bra{d_{-k,0}^{(\lambda)}} v\tau_z \ket{d_{k,0}^{(\lambda)\dagger}}
    }^2}{(2E_{\rm bulk}(k,t=0))^4}
    +\frac{\abs{
    \bra{d_{-k,T}^{(\lambda)}} v\tau_z \ket{d_{k,T}^{(\lambda)\dagger}}
    }^2}{(2E_{\rm bulk}(k,t=T))^4} \nonumber \\
    &&-2\frac{\bra{d_{-k,0}^{(\lambda)}} v\tau_z \ket{d_{k,0}^{(\lambda)\dagger}}}{(2E_{\rm bulk}(k,t=0))}
    \frac{\bra{d_{-k,T}^{(\lambda)}} v\tau_z \ket{d_{k,T}^{(\lambda)\dagger}}}{(2E_{\rm bulk}(k,t=T))}\cos(\Delta\theta_{\bf n}),
\end{eqnarray}
For small changes in the chemical potential, we can approximate
\begin{equation}
    \bra{d_{-k,0}^{(\lambda)}} \tau_z \ket{d_{k,0}^{(\lambda)\dagger}}
    \approx \bra{d_{-k,T}^{(\lambda)}} \tau_z \ket{d_{k,T}^{(\lambda)\dagger}}
\end{equation}
and $E_{\rm bulk}(k,t=0) = E_{\rm bulk}(k,t=T) = E_{\rm bulk}(k)$. 
We then see that $L_{\rm even}$ oscillates with $v$ sinusoidally and has maximum amplitude
\begin{equation}
    \abs{\beta_k^{(\lambda)}}^2 \approx \frac{v^2\abs{
    \bra{d_{-k,0}^{(\lambda)}} \tau_z \ket{d_{k,0}^{(\lambda)\dagger}}
    }^2}{4E_{\rm bulk}(k)^4}
\end{equation}
By using the exact solution for the Kitaev chain, we obtain
\begin{equation}
    \abs{\beta_k^{(\lambda)}}^2 \approx \frac{v^2 (\delta_k (\varepsilon_k-E_{\rm bulk}(k)))^2} 
    {(\delta_k^2+(\varepsilon_k-E_{\rm bulk}(k))^2)  E_{\rm bulk}(k)^4}
\end{equation}
Now for $\mu = 0$ and $w = \Delta = 1/2$, we have
$\delta_k = \sin{k}$, $\varepsilon_k = \cos{k}$ and 
$E_{\rm bulk}(k) = 1$. This gives us $\abs{\beta_k^{(\lambda)}}^2 = v^2\sin^2{k}/4$.
Finally, from Eq.~\eqref{eq:leventobeta}, we get
\begin{equation}
    L_{\rm even} \approx \tilde{L}_{\rm even, near-ad} \equiv \frac{Nv^2}{8\pi}\int_{-\pi}^{\pi}\sin^2{k}dk = \frac{Nv^2}{8}. 
    \label{eqn:even_adiabatic_approx}
\end{equation}

At sufficiently low but finite $\mu_{\rm fin}$ in the near-adiabatic regime, the leakage $L_{\rm even}$ is expected to oscillate between $0$ and $\tilde{L}_{\rm even, near-ad}$ as a function of $v$. Note that this expression is remarkably independent of $\mu_{\rm fin}$. In our numerics we observe good agreement with this expression and $L_{\rm even}$ as shown in Fig. \ref{fig:near_adibatic_osc}, particularly at sufficiently low $\mu_{\rm fin} = 0.01$. Furthermore, in the near-adiabatic regime, as with the sudden regime, we find in our numerics that at sufficiently long Kitaev chain lengths $N$, that $L_{\text{even}}$ grows linearly in $N$, whereas $L_{\text{odd}}$ is constant in $N$. This is exemplified in Fig. \ref{fig:near_adiabatic_scaling} for various final chemical potentials ($\mu_{\text{fin}}$) that are deep in the topological phase $\mu_{\text{fin}} \ll 2|w|$ and in the near-adiabatic regime ($v = 10^{-3}$). 

\begin{figure}[h!]
\includegraphics[width=17cm]{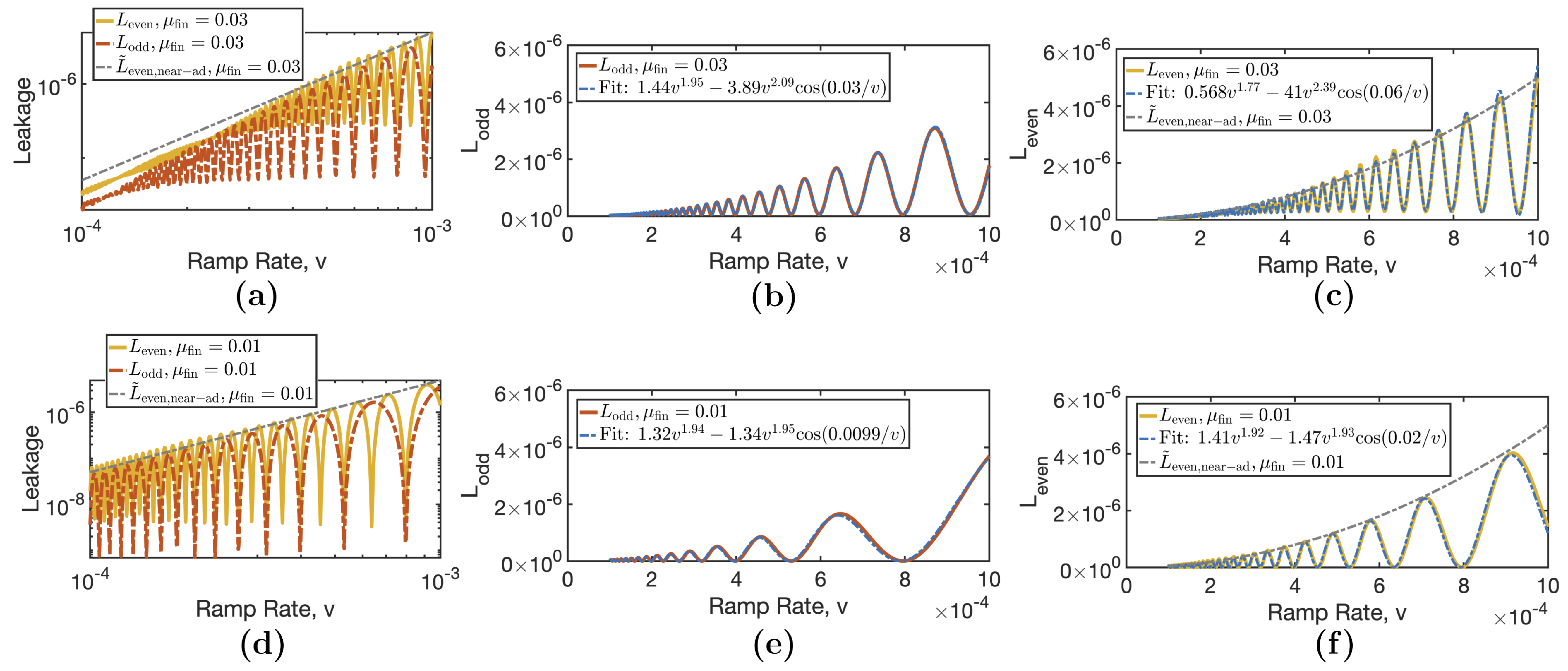}
\caption{Near-adiabatic regime leakage into the sectors with odd ($L_\text{odd}$) and even ($L_{\text{even}}$) numbers of quasiparticles (QPs) versus chemical potential ramp rate ($v$) for a Kitaev-tetron qubit with initial chemical potential $\mu_\text{in} = 0$, with hopping ($w$) and pairing ($\Delta$) of $w = \Delta = 0.5$, and final chemical potential $\mu_\text{fin} = 0.03$ (a-c) and $\mu_{\text{fin}} = 0.01$ (d-f). Panels (a,~d) show the leakages on a log-log scale for chemical potential ramp rates $v \in [10^{-4},\, 10^{-3}]$. Panels (b,~e) and (c,~d) respectively show $L_\text{odd}$ and $L_\text{even}$ on a linear scale for $v \in [10^{-4},\, 10^{-3}]$. The grey dashed line in panels (a,~d) and (c,~d) shows $\tilde{L}_{\text{even, near-ad}}$ (see Eq. \ref{eqn:even_adiabatic_approx}) which is the predicted maxima in $L_{\text{even}}$ at small $\mu_{\rm fin}$, which is in close agreement with the numerics particularly at the smaller $\mu_{\rm fin} = 0.01$. The leakages in each panel (b,~c,~e,~f) are fit to $L = k_1 v^m_1 + k_2^{m_2} \text{cos}(\omega /v)$, which is the the expected form for leakage into a single level as given in Eq.~(\ref{eqn:originaladiabatic}),  over the reduced domain $v \in [4 \times 10^{-4},\, 10^{-3}]$. A reduced domain was used since on larger intervals of $v$, the leakage to the slightly dispersed leakage levels causes beats to arise and so $L_{\text{even/odd}}$ deviates slightly from Eq.~(\ref{eqn:originaladiabatic}). Note that the oscillations in $L_{\text{odd}}$ with frequencies $\omega_{\text{odd}}(\mu_{\rm final} = 0.03) = 0.030$ and $\omega_{\text{odd}}(\mu_{\rm final} = 0.01) = 0.099$ are in agreement with Eq.~(\ref{eqn:dynamic_phase}) and Eq.~(\ref{eqn:originaladiabatic}). Furthermore, the oscillations of $L_{\text{even}}$ are double the frequency at $\omega_{\text{even}}(\mu_{\rm final} = 0.03) = 0.060$ and $\omega_{\text{even}}(\mu_{\rm final} = 0.01) = 0.020$ as expected. In further agreement Eq.~(\ref{eqn:originaladiabatic}), $m_{1,\text{odd}} \approx m_{2, \text{odd}} \approx 2$ for both values of $\mu_{\rm final}$. Whereas $m_{1,\text{odd}}$ and  $m_{2, \text{odd}}$ differ slightly from $2$ due to the interference of the leakage levels, which causes beats that are particularly pronounced at $\mu_{\rm final} = 0.03$ as shown in panel (a).}
\label{fig:near_adibatic_osc}
\end{figure}

\begin{figure}[h!]
\includegraphics[width=17cm]{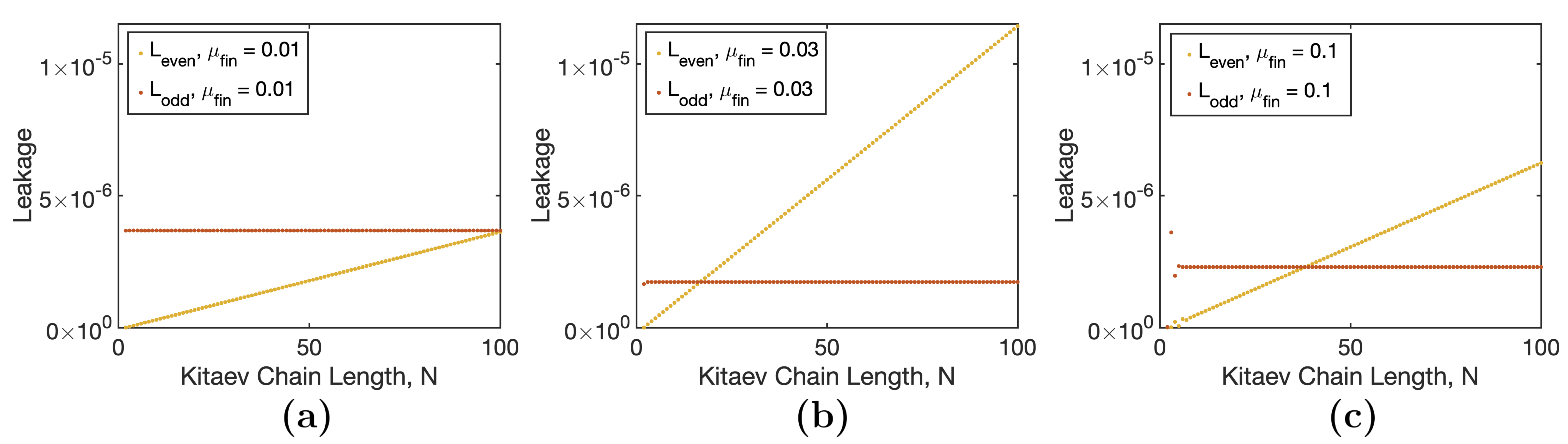}
\caption{Scaling of leakages into states with even ($L_{\text{even}}$) and odd ($L_\text{odd}$) numbers of quasiparticles at a chemical potential ramp rate of $v = 10^{-3}$ (which is in the near-adiabatic regime), for a Kitaev-tetron qubit with initial chemical potential $\mu_{\text{in}} = 0$, and with hopping ($w$) and pairing ($\Delta$) of $w = \Delta = 0.5$. Each panel presents the leakage versus the Kitaev chain lengths $N$ for all $N$ on the interval $N \in [2, 100]$. The panels respectively consider final chemical potentials of (a) $\mu_\text{fin} = 0.01$, (b) $\mu_\text{fin} = 0.03$, and (c) $\mu_\text{fin} = 0.1$. For sufficiently large chains $L_{\text{even}}$ is linear in $N$ and $L_{\text{odd}}$ is constant, as also observed in the sudden limit (as shown in Fig. \ref{fig:sudden_leakage_scaling}) and at intermediate ramp rates between the near-adiabatic and sudden regimes (as shown in Fig. 2 in the main text).}
\label{fig:near_adiabatic_scaling}
\end{figure}

\section{Derivation of Probability of Absorption of Quasiparticles at Opposite Nanowire Ends}
\label{sec:app_e}

This section of appendices presents more details concerning the argument presented in the discussion section of the main text, which claims that when pairs of quasiparticles (QPs) are excited in the bulk, a nonzero fraction of these pairs will induce qubit errors.  As stated in the main text, these arguments are based on: (1) the rate at which QP pairs decay back into the superconducting ground state is small \cite{karzig_2021_phys_rev_let_126}; (2) a qubit error occurs if two quasiparticles are created and then absorbed by MZMs at opposite ends of a chain \cite{knapp_2018_PRB_97_12, Alase_2024_Phys_Rev_Res_6}; and (3) the motion of quasiparticles in long nanowires is diffusive \cite{Mayer_2020_NatureComm_11_12, karzig_2021_phys_rev_let_126}.

Here we show that if each QP undergoes diffusive motion and is absorbed by an MZM as soon as it reaches an end of the chain, then a nonzero fraction of the excited QP pairs in the bulk cause qubit errors.  The fraction of QP pairs that cause qubit errors is $1/3$ in the limit of long chains.

We first consider two QPs that both start at a point $x$ in the domain
$[0,L]$ and then undergo independent unbiased random walks until they both have reached a chain end and been absorbed by an MZM \cite{karzig_2021_phys_rev_let_126}. The convention used here for numbering the sites, which differs slightly from that in the main text, is used because the resulting expressions are particularly simple.
The probability that two particles that start at a point $x$ are absorbed at different ends, $P_{\text{different}}(x)$,  is
\begin{equation}
P_{\text{different}}(x) = 2P_0(x)P_L(x)~ 
= 1-P_0^2(x)-P_L^2(x)~,
\label{eq:P_different}
\end{equation}
where $P_0(x)$ and $P_L(x)$ are the probabilities that a walker starting out at point $x$ is absorbed at $0$ and $L$, respectively.
The validity of Eq.~(\ref{eq:P_different}) can be seen in two ways, both by noting it is the sum of the probabilities that the first QP is absorbed at $0$ and the second is absorbed at $L$ and the probability the first QP is absorbed at $L$ and the second is absorbed at $0$, or as one minus the probability that both QPs are absorbed at the same end of the nanowire.
Ref.~\cite{Doyle_1984_AMS_22} shows that 
\begin{equation}
P_L(x) = 1-P_0(x) = \frac{x}{L}~.
\label{eq:P_L}
\end{equation}
The derivation of Eq.~(\ref{eq:P_L}) from Ref.~\cite{Doyle_1984_AMS_22} will be presented below for completeness, but first we use Eq.~(\ref{eq:P_L}) to compute the probability that two QPs created at the same site $x$, which is distributed uniformly in the bulk, are absorbed at opposite ends of the wire.  If the probability of exciting a QP pair is uniform across the chain, then the average probability that two QPs emitted at the same point are absorbed at opposite ends, $P_{\text{average}}$, is
\begin{equation}
P_{\text{average}} = \frac{1}{L+1}\sum_{x=0}^{x=L}
P_{\text{different}}(x) = \frac{1}{3}\left ( 1-\frac{1}{L} \right )~.
\label{eq:difference_integrated}
\end{equation}
In the limit $L \rightarrow\infty$, $P_\text{{average}} \rightarrow 1/3$, as quoted in the discussion section of the main text. Now, for completeness, we summarize the derivation of Eq.~(\ref{eq:P_L}) from Ref.~\cite{Doyle_1984_AMS_22}.
We note that the walker that starts at $x=0$ at time $0$ is immediately absorbed at $x=0$, so $P_L(0)=0$.
Similarly, a walker that starts at $x=L$ is immediately absorbed at $x=L$, so $P_L(L)=1$.
Now consider an $x$ in the chain interior.
After one step of the walk, the particle that starts out at $x$ has equal probability of being at $x-1$ or at $x+1$,
so $P_L(x) = (P_L(x-1)+P_L(x+1))/2$.
It is straightforward to plug in and verify that $P_L(x)=x/L$ solves the equations
\begin{eqnarray}
P_L(0) &=& 0 \nonumber \\
P_L(L) &=& 1 \nonumber \\
P_L(x) &=& \frac{1}{2} \left ( P_L(x-1) + P_L(x+1) \right )~  {\rm ~for~ } 0<x<L~,
\end{eqnarray}
and it is shown in Ref.~\cite{Doyle_1984_AMS_22} that this solution is unique.

\twocolumngrid
\bibliography{QPP_lib}

\end{document}